\documentclass[leqno, fleqn, 12pt, preprint]{elsarticle}
\usepackage{MathSpad,makeidx,ifthen,latexsym}
\usepackage{amssymb, amsfonts, latexsym,graphs,graphicx,url,amsmath}

\def\MPsf#1{\mathsf{#1}}
\def\@mathbf#1{#1}
\def\mathit#1{#1}

\begin{document}
\title{On Euclid's Algorithm and Elementary Number Theory}

\author{Roland Backhouse}
\ead{rcb@cs.nott.ac.uk}

\author{Jo\~ao F. Ferreira\fnref{funded-fct}}
\ead{joao@joaoff.com}

\fntext[funded-fct]{Funded by
Funda\c{c}\~ao para a Ci\^encia e a Tecnologia (Portugal) under grant
SFRH/BD/24269/2005}

\address{School of Computer Science, University of Nottingham, Nottingham, NG8 1BB, England}
\date{\today}

\begin{abstract}
Algorithms can be used to prove and to discover new theorems. This paper shows how algorithmic skills
in general, and the notion of invariance in particular, can be used to derive many results from Euclid's
algorithm. We illustrate how to use the algorithm as a verification interface (i.e., how to verify
theorems) and as a construction interface (i.e., how to investigate and derive new theorems). 

The theorems that we verify are well-known and most of them are included in standard number-theory
books. The new results concern distributivity properties of the greatest common divisor and a new
algorithm for efficiently enumerating the positive rationals in two different ways. One way is known
and is due to Moshe Newman. The second  is new and corresponds to a deforestation  of the Stern-Brocot
tree of rationals.   We show that both enumerations stem from the same simple algorithm.  In this way,
we construct a Stern-Brocot enumeration algorithm with the same time and space complexity as
Newman's algorithm.  A short review of the original papers by Stern and Brocot is also included.
\end{abstract}


\maketitle


\section{Introduction}\label{Introduction}
An algorithm is a sequence of instructions that can be systematically executed in
the solution of a given problem. Algorithms have been studied and
developed since the beginning of civilisation, but, over the last 50 years,
the unprecedented scale of programming problems and the consequent demands on precision and concision 
have made computer scientists hone their algorithmic problem-solving skills to a  fine degree.

Even so, and although much of mathematics is algorithmic in nature, the 
skills needed to formulate and solve algorithmic problems do not 
form an integral part of contemporary mathematics education; also, the teaching 
of computer-related topics at pre-university level focuses on enabling 
students to be  effective users of information technology, rather 
than equip them with the skills to develop new applications or to solve
new problems.

A blatant example is the conventional  treatment of Euclid's algorithm to compute the greatest common divisor (gcd)
of two positive natural numbers, the oldest nontrivial algorithm
that involves iteration and that has not been superseded by algebraic methods. (For a modern paraphrase of
Euclid's original statement, see \cite[pp. 335--336]{DEK-TAOCP2}.)
Most books on number theory include Euclid's algorithm,  but rarely use the algorithm
directly to reason about properties of numbers.  Moreover, the presentation of the algorithm in such books has
benefited little from the advances that have been made in our understanding of the basic principles of
algorithm development.  In an article such as this one, it is of course not the place to rewrite mathematics
textbooks.  Nevertheless, our goal in this paper is to demonstrate how a focus on algorithmic method can
enrich and re-invigorate the teaching of mathematics.  We use Euclid's algorithm to derive  both old and 
well-known,  and  new and previously unknown, 
properties of the greatest common divisor and rational numbers.  The leitmotiv is the notion of a loop invariant
--- how it can be used as a verification 
interface (i.e., how to verify theorems) and as a construction interface (i.e., how to investigate and derive 
new theorems).  

We begin the paper in section \ref{Divisibility Theory} with basic  properties of the division  relation and the
construction of  Euclid's algorithm from its formal specification.    In contrast to  standard presentations of the
algorithm, which  typically assume the existence of the gcd operator with specific algebraic properties, our
derivation gives a constructive proof of the existence of an infimum operator in the division ordering of
natural numbers. 

 The focus of section \ref{Euclid's Algorithm as a Verification Interface} is the systematic use of 
 invariant properties of Euclid's  algorithm to verify  known identities.   Section 
\ref{Euclid's Algorithm as a Construction Interface}, on the other hand, shows how to use the algorithm to derive new
results related with the greatest common divisor:  we calculate  sufficient conditions for a
natural-valued function\footnote{We call a function natural-valued if its range is the set of natural numbers.} to distribute over the greatest common divisor, and we derive an efficient algorithm
to enumerate the positive rational numbers in two different ways.  

Although the identities in section \ref{Euclid's Algorithm as a Verification Interface} are well-known, we believe
that our derivations improve considerably on standard presentations.  One example is the proof 
that the greatest common divisor of two numbers is a linear combination of the numbers; by the
simple device of introducing matrix arithmetic into Euclid's algorithm, it suffices to observe that matrix
multiplication is associative in order to prove the theorem.   This exemplifies the gains in 
our problem-solving skills that can be achieved  by the right combination of precision and concision.  The
introduction of matrix arithmetic at this early stage was also what enabled us to derive a previously unknown
algorithm to enumerate the rationals in so-called Stern-Brocot order (see section 
\ref{Euclid's Algorithm as a Construction Interface}), which is the primary  novel result (as opposed to method)
in this paper.

Included in the appendix is a brief summary of the work of Stern and Brocot, the 19th
century authors after whom the Stern-Brocot tree is named. It is interesting to review their work, particularly
that of Brocot, because it is clearly motivated by practical, algorithmic problems.  The review of Stern's paper
is included in order to resolve recent misunderstandings about the origin of the Eisenstein-Stern and Stern-Brocot
enumerations of the rationals.

\section{Divisibility Theory}\label{Divisibility Theory}

Division is one of the most important concepts in number theory. 
This section begins with a short,  basic account of the division 
relation.  We observe that division is a partial ordering on the natural numbers and pose the question
whether the infimum, in the division ordering, of any pair of numbers exists.     The algorithm we know
as Euclid's gcd algorithm is then derived in order to give a positive (constructive) answer to this question.

\subsection{Division Relation}\label{Division Relation}
The division relation, here denoted by an infix ``${\setminus}$'' symbol, is the relation on integers defined 
to be the converse of the ``is-a-multiple-of'' relation\footnote{The square so-called ``everywhere''  brackets are used
to indicate that a boolean statement is ``everywhere'' true.   That is, the statement has the value \textsf{true} for
all instantiations of its  free variables.  Such statements are often called ``facts'',  or ``laws'', or ``theorems''.

When using the everywhere brackets, the domain of the free
variables has to be made clear.  This is particularly important here because sometimes the domain of a
variable is the integers and sometimes it is the natural numbers. Usually, we rely on a convention for
naming the variables, but sometimes we preface a law with a reminder of the domain.

Also, we use a systematic notation for quantified expressions which has the form $\langle\bigoplus\mathit{bv\/}{:}\ms{2}\mathit{range\/}{:}\ms{2}\mathit{term\/}\rangle$.
There are five components to the notation, which we explain in turn. The first component is the quantifier,
in this case $\bigoplus$. The second component is the dummy $\mathit{bv\/}$. The third component is the range of the dummy,
a boolean-valued expression that determines a set of values of the dummy. The fourth component is
the term. The final component of the notation is the angle brackets; they serve to delimit the scope of the dummy.
For more details, see \cite[chapter 11]{PC03} and \cite[chapter 8]{GS93a}.

The symbol \setms{0.2em}${\equiv}$ denotes boolean equality. In continued expressions ${\equiv}$ is read associatively and ${=}$ is read conjunctionally.
For example, $p\ms{1}{\equiv}\ms{1}q\ms{1}{\equiv}\ms{1}r$ is evaluated associatively ---i.e. as $(p\ms{1}{\equiv}\ms{1}q)\ms{1}{\equiv}\ms{1}r$ or $p\ms{1}{\equiv}\ms{1}(q\ms{1}{\equiv}\ms{1}r)$, whichever is most convenient--- whereas
$p\ms{1}{=}\ms{1}q\ms{1}{=}\ms{1}r$ is evaluated conjunctionally--- i.e. $p\ms{1}{=}\ms{1}q$ and $q\ms{1}{=}\ms{1}r$.}:
\begin{displaymath}[\ms{3}\ms{3}m{\setminus}n\ms{3}{\equiv}\ms{3}\langle\exists{}k\ms{2}{:}\ms{2}k{\in}\MPInt\ms{2}{:}\ms{2}n\ms{1}{=}\ms{1}k{\times}m\rangle\ms{3}]\mbox{\ \ \ .}\end{displaymath}In words, an integer $m$ divides an integer $n$ (or $n$ is divisible by $m$) if there exists some integer $k$ such
that $n\ms{1}{=}\ms{1}k{\times}m$. In that case, we say that $m$ is a divisor of $n$ and that $n$ is a multiple of $m$.

The division relation plays a prominent role in number theory.  So, we start by presenting some of its 
basic properties and their relation to addition and multiplication. First, it is reflexive because multiplication 
has a unit (i.e., $m\ms{1}{=}\ms{1}1{\times}m$) and it is transitive, since multiplication is associative.
It is also (almost) preserved by linear combination because multiplication distributes over addition:
\begin{equation}\label{prop-linear-comb}
[\ms{3}k{\setminus}x\ms{2}{\wedge}\ms{2}k{\setminus}y\ms{4}{\equiv}\ms{4}k{\setminus}(x\ms{1}{+}\ms{1}a{\times}y)\ms{2}{\wedge}\ms{2}k{\setminus}y\ms{3}]\mbox{\ \ \ .}
\end{equation}(We leave the reader to verify this law; take care to note the use of the distributivity of multiplication
over addition in its proof.)
Reflexivity and transitivity make  division a \emph{preorder} on the
integers.    It is not anti-symmetric but the  numbers equivalent under the preordering are given
by\begin{displaymath}[\ms{3}m{\setminus}n\ms{2}{\wedge}\ms{2}n{\setminus}m\ms{4}{\equiv}\ms{4}\mathsf{abs}{.}m\ms{1}{=}\ms{1}\mathsf{abs}{.}n\ms{3}]\mbox{\ \ \ ,}\end{displaymath}where $\mathsf{abs}$ is the absolute value function and the infix dot denotes function application.
Each equivalence class thus consists of a natural number and its negation.  
If the division relation is restricted to   natural numbers,  division becomes anti-symmetric,
since $\mathsf{abs}$ is the identity function on natural numbers.
This means that,  restricted to  the natural numbers,  division  is a
\emph{partial} order  with $0$ as the greatest element  and $1$ as the smallest element.


\subsubsection{Infimum in the Division Ordering}\label{Division-Infimum}
The first question that we consider is whether two arbitrary  natural numbers $m$ and $n$  have an infimum  
in the division ordering.   That is, can we solve the following equation\footnote{Unless indicated otherwise, the domain of
all variables is $\MPNat$, the set of  natural numbers.  Note that we include $0$ in $\MPNat$. The notation $x{:}{:}\ms{1}E$ means that $x$ is
the unknown and the other free variables are parameters of the equation $E$.}?\begin{equation}\label{nabla}
x{:}{:}\ms{9}\left\langle\forall{}k\ms{5}{:}{:}\ms{5}k{\setminus}m\ms{2}{\wedge}\ms{2}k{\setminus}n\ms{5}{\equiv}\ms{5}k{\setminus}x\right\rangle~~.
\end{equation}The answer  is not immediately obvious because the division ordering is partial.  (With respect to a total
ordering, the infimum  of two numbers is their minimum;  it is thus equal to one of them and can be
easily computed by a case analysis.)

If a solution to (\ref{nabla}) exists, it is unique (because the division relation  on natural numbers is reflexive and
anti-symmetric).   When it  does have a solution, we denote it by $m\protect\raisebox{0.8mm}{$\bigtriangledown$}n$.  
That is, provided it can be established that (\ref{nabla})
has a solution,
\begin{equation}\label{nabla0}
[\ms{3}k{\setminus}m\ms{2}{\wedge}\ms{2}k{\setminus}n\ms{6}{\equiv}\ms{6}k\ms{1}{\setminus}\ms{1}(m\protect\raisebox{0.8mm}{$\bigtriangledown$}n)\ms{3}]\mbox{\ \ \ .}
\end{equation}Because conjunction is idempotent,\begin{displaymath}[\ms{3}k{\setminus}m\ms{2}{\wedge}\ms{2}k{\setminus}m\ms{6}{\equiv}\ms{6}k{\setminus}m\ms{3}]\mbox{\ \ \ .}\end{displaymath}That is, $m$ solves (\ref{nabla}) when $m$ and $n$ are equal.  Also,    because    $[\ms{3}k{\setminus}0\ms{3}]$,  \begin{displaymath}[\ms{3}k{\setminus}m\ms{2}{\wedge}\ms{2}k{\setminus}0\ms{6}{\equiv}\ms{6}k{\setminus}m\ms{3}]\mbox{\ \ \ .}\end{displaymath}That is,   $m$ solves (\ref{nabla}) when  $n$ is $0.$  So,   $m\ms{0}\raisebox{0.8mm}{\ensuremath{\bigtriangledown}} m$ exists as does $m\ms{0}\raisebox{0.8mm}{\ensuremath{\bigtriangledown}} 0$, and both equal $m$:\begin{equation}\label{nabla1}
[\ms{2}m\protect\raisebox{0.8mm}{$\bigtriangledown$}m\ms{3}{=}\ms{3}m\protect\raisebox{0.8mm}{$\bigtriangledown$}0\ms{3}{=}\ms{3}m\ms{2}]\mbox{\ \ \ .}
\end{equation}
Other properties that are easy to establish by exploiting the algebraic properties of conjunction are, first, $\protect\raisebox{0.8mm}{$\bigtriangledown$}$
is symmetric (because conjunction is symmetric)\begin{equation}\label{nabla2}
[\ms{2}m\ms{0}\raisebox{0.8mm}{\ensuremath{\bigtriangledown}} n\ms{3}{=}\ms{3}n\protect\raisebox{0.8mm}{$\bigtriangledown$}m\ms{2}]\mbox{\ \ \ ,}
\end{equation}and, second,  $\protect\raisebox{0.8mm}{$\bigtriangledown$}$ is associative (because conjunction is associative)\begin{equation}\label{nabla3}
[\ms{2}(m\ms{0}\raisebox{0.8mm}{\ensuremath{\bigtriangledown}} n)\ms{0}\raisebox{0.8mm}{\ensuremath{\bigtriangledown}} p\ms{2}{=}\ms{2}m\protect\raisebox{0.8mm}{$\bigtriangledown$}(n\protect\raisebox{0.8mm}{$\bigtriangledown$}p)\ms{2}]\mbox{\ \ \ .}
\end{equation} Note that we choose infix notation for $\protect\raisebox{0.8mm}{$\bigtriangledown$}$, since it allows us to write
$m\protect\raisebox{0.8mm}{$\bigtriangledown$}n\protect\raisebox{0.8mm}{$\bigtriangledown$}p$ without having to choose between $(m\protect\raisebox{0.8mm}{$\bigtriangledown$}n)\protect\raisebox{0.8mm}{$\bigtriangledown$}p$ or $m\protect\raisebox{0.8mm}{$\bigtriangledown$}(n\protect\raisebox{0.8mm}{$\bigtriangledown$}p)$.

The final property of $\protect\raisebox{0.8mm}{$\bigtriangledown$}$ that we  deduce from (\ref{nabla0}) is obtained by exploiting (\ref{prop-linear-comb}), with $x$ and $y$
replaced by $m$ and $n$, respectively :\begin{equation}\label{nabla-lin-comb}
[\ms{3}(m\ms{1}{+}\ms{1}a{\times}n)\protect\raisebox{0.8mm}{$\bigtriangledown$}n\ms{5}{=}\ms{5}m\protect\raisebox{0.8mm}{$\bigtriangledown$}n\ms{3}]\mbox{\ \ \ .}
\end{equation}
\subsection{Constructing Euclid's Algorithm}\label{Constructing Euclid's Algorithm}
At this stage in our analysis, properties (\ref{nabla2}),  (\ref{nabla3}) and (\ref{nabla-lin-comb}) 
assume that equation (\ref{nabla}) has a
solution in the appropriate cases.  For instance,  (\ref{nabla2}) means that, if (\ref{nabla}) has a solution for
certain  natural numbers $m$ and $n$, it also has a solution when the values of $m$ and $n$ are
interchanged.  

In view of properties (\ref{nabla1}) and  (\ref{nabla2}), it remains to show that  (\ref{nabla}) has a solution when both
$m$ and $n$ are strictly positive and unequal. We do this  by providing an algorithm that computes the
solution.  
Equation  (\ref{nabla}) does not directly suggest any algorithm, but the germ of an algorithm is suggested 
by observing that it is equivalent to\begin{equation}\label{gcd6}
x,y{:}{:}\ms{9}x\ms{1}{=}\ms{1}y\ms{5}{\wedge}\ms{5}\langle\forall{}k{:}{:}\ms{4}k{\setminus}m\ms{2}{\wedge}\ms{2}k{\setminus}n\ms{4}{\equiv}\ms{4}k{\setminus}x\ms{2}{\wedge}\ms{2}k{\setminus}y\rangle\mbox{\ \ \ .}
\end{equation}This new shape strongly suggests an algorithm that, initially, establishes the truth of\begin{displaymath}\langle\forall{}k{:}{:}\ms{4}k{\setminus}m\ms{2}{\wedge}\ms{2}k{\setminus}n\ms{4}{\equiv}\ms{4}k{\setminus}x\ms{2}{\wedge}\ms{2}k{\setminus}y\rangle\end{displaymath} ---which is trivially achieved by the assignment $x{,}y:=m{,}n$---  and then,  reduces $x$ and $y$  in such a way
that the property  is kept invariant whilst making progress to  a  state satisfying $x\ms{1}{=}\ms{1}y$.  When such a
state is reached, we have found a solution to the equation (\ref{gcd6}), and the value of $x$  (or $y$ since they are
equal) is a solution of (\ref{nabla}).  Thus, the structure of the algorithm we are trying to develop is as 
follows\footnote{We use the Guarded Command Language (GCL), a very simple programming language
with just four programming constructs---assignment, sequential composition, conditionals, and loops. The GCL was
introduced by Dijkstra \cite{Dij75}. The statement $\MPsf{do}~S~\MPsf{od}$ is a loop that executes $S$ repeatedly while at least one of $S$'s 
guards is true. Expressions in curly brackets are assertions.}:
\begin{mpdisplay}{0.2em}{5.1mm}{2mm}{2}
	\push$\{~~$\=\+ $0\ms{1}{<}\ms{1}m\ms{3}{\wedge}\ms{3}0\ms{1}{<}\ms{1}n$\-$ ~~\}$\pop\\
	$x\ms{1}{,}\ms{1}y\ms{1}:=\ms{1}m\ms{1}{,}\ms{1}n\ms{2};$\\
	\push$\{~~$\=\+\push\textbf{Invariant}:$~~~~$\=\+$\langle\forall{}k{:}{:}\ms{4}k{\setminus}m\ms{2}{\wedge}\ms{2}k{\setminus}n\ms{4}{\equiv}\ms{4}k{\setminus}x\ms{2}{\wedge}\ms{2}k{\setminus}y\rangle$\-\pop\-$ ~~\}$\pop\\
	\push$\MPsf{do}\ms{3}$\=\+$x\ms{1}{\neq}\ms{1}y\ms{2}\MPsf{\rightarrow}\ms{2}x\ms{1}{,}\ms{1}y\ms{1}:=\ms{1}A\ms{1}{,}\ms{1}B$\-\\
	$\MPsf{od}$\pop\\
	\push$\{~~$\=\+$x\ms{1}{=}\ms{1}y\ms{5}{\wedge}\ms{5}\langle\forall{}k{:}{:}\ms{4}k{\setminus}m\ms{2}{\wedge}\ms{2}k{\setminus}n\ms{4}{\equiv}\ms{4}k{\setminus}x\ms{2}{\wedge}\ms{2}k{\setminus}y\rangle$\-$ ~~\}$\pop
\end{mpdisplay}
Now we only have to define $A$ and $B$ in such a way that the assignment in the loop body leads to a state
where $x\ms{1}{=}\ms{1}y$ is satisfied while maintaining the invariant.   Exploiting the transitivity of equality, the invariant
is maintained  by choosing $A$ and $B$ so that 
\begin{equation}\label{gcdinv}
\langle\forall{}k{:}{:}\ms{4}k{\setminus}x\ms{2}{\wedge}\ms{2}k{\setminus}y\ms{4}{\equiv}\ms{4}k{\setminus}A\ms{2}{\wedge}\ms{2}k{\setminus}B\rangle~~.
\end{equation}To ensure that we are making progress towards the termination
condition, we have to define a \emph{bound function}, which is
a natural-valued function of the  variables $x$ and $y$ that measures
the size of the problem to be solved.  A guarantee that the value of
such a bound function is always decreased at each iteration is a guarantee
that the number of times the loop body is executed is at most the initial value of 
the bound function. The definition of the bound function depends  on
the assignments we choose for $A$ and $B.$

At this point, we need to exploit properties specific to division. 
(Refer back to section \ref{Division Relation} for a discussion of some of the properties.)
Inspecting the shape of (\ref{gcdinv}), we see that it is similar
to the shape  of property (\ref{prop-linear-comb}). This suggests that
we can use (\ref{prop-linear-comb}), and in fact, considering this property,
we have the corollary:
\begin{equation}\label{prop-inv1}
[\ms{3}k{\setminus}x\ms{2}{\wedge}\ms{2}k{\setminus}y\ms{4}{\equiv}\ms{4}k{\setminus}(x{-}y)\ms{2}{\wedge}\ms{2}k{\setminus}y\ms{3}]\mbox{\ \ \ .}
\end{equation}The relevance of  this corollary is that our invariant is preserved by the
assignment $x\ms{1}:=\ms{1}x{-}y$ (leaving the value of $y$ unchanged).   (Compare (\ref{prop-inv1}) with (\ref{gcdinv}).)
Note that this also reduces the value of $x$ when $y$
is positive.  This suggests that we strengthen the invariant by requiring that $x$ and $y$ remain positive; the
assignment $x\ms{1}:=\ms{1}x{-}y$ is executed when $x$ is greater than $y$ and, symmetrically, the assignment $y\ms{1}:=\ms{1}y{-}x$ is
executed when $y$ is greater than $x$.  As bound function we can take $x{+}y$.  
The  algorithm becomes

\begin{mpdisplay}{0.2em}{5.1mm}{2mm}{2}
	\push$\{~~$\=\+$0\ms{1}{<}\ms{1}m\ms{3}{\wedge}\ms{3}0\ms{1}{<}\ms{1}n$\-$ ~~\}$\pop\\
	$x\ms{1}{,}\ms{1}y\ms{1}:=\ms{1}m\ms{1}{,}\ms{1}n\ms{2};$\\
	\push$\{~~$\=\+\push\textbf{Invariant}:$~~~~$\=\+$0{<}x\ms{3}{\wedge}\ms{3}0{<}y\ms{3}{\wedge}\ms{3}\langle\forall{}k{:}{:}\ms{3}k{\setminus}m\ms{2}{\wedge}\ms{2}k{\setminus}n\ms{4}{\equiv}\ms{4}k{\setminus}x\ms{2}{\wedge}\ms{2}k{\setminus}y\rangle$\-\pop\\
	\textbf{Bound function}: $x{+}y$\-$ ~~\}$\pop\\
	\push$\MPsf{do}\ms{2}$\=\+\push$x\ms{1}{\neq}\ms{1}y~~\MPsf{\rightarrow}$\\
	$~~~~$\=\+$~$\=\+\push$\MPsf{if}\ms{5}$\=\+$y\ms{1}{<}\ms{1}x\ms{2}\MPsf{\rightarrow}\ms{2}x\ms{1}:=\ms{1}x{-}y$\push\-\\
	$\MPsf{\Box}$	\>\+\pop$x\ms{1}{<}\ms{1}y\ms{2}\MPsf{\rightarrow}\ms{2}y\ms{1}:=\ms{1}y{-}x$\-\\
	$\MPsf{fi}$\pop\-\-\pop\-\\
	$\MPsf{od}$\pop\\
	\push$\{~~$\=\+$0{<}x\ms{4}{\wedge}\ms{4}0{<}y\ms{4}{\wedge}\ms{4}x\ms{1}{=}\ms{1}y\ms{4}{\wedge}\ms{4}\langle\forall{}k{:}{:}\ms{3}k{\setminus}m\ms{2}{\wedge}\ms{2}k{\setminus}n\ms{4}{\equiv}\ms{4}k{\setminus}x\ms{2}{\wedge}\ms{2}k{\setminus}y\rangle$\-$ ~~\}$\pop
\end{mpdisplay}
(We leave the reader to perform the standard steps used to verify the correctness of the algorithm.)
Finally, since \begin{displaymath}(x\ms{1}{<}\ms{1}y\ms{2}{\vee}\ms{2}y\ms{1}{<}\ms{1}x)\ms{4}{\equiv}\ms{4}x\ms{1}{\neq}\ms{1}y\mbox{\ \ \ ,}\end{displaymath}we can safely remove the outer guard and
simplify the algorithm, as shown below.

\begin{mpdisplay}{0.2em}{5.1mm}{2mm}{2}
	\push$\{~~$\=\+$0\ms{1}{<}\ms{1}m\ms{3}{\wedge}\ms{3}0\ms{1}{<}\ms{1}n$\-$ ~~\}$\pop\\
	$x\ms{1}{,}\ms{1}y\ms{1}:=\ms{1}m\ms{1}{,}\ms{1}n\ms{2};$\\
	\push$\{~~$\=\+\push\textbf{Invariant}:$~~~~$\=\+$0{<}x\ms{4}{\wedge}\ms{4}0{<}y\ms{4}{\wedge}\ms{4}\langle\forall{}k{:}{:}\ms{3}k{\setminus}m\ms{2}{\wedge}\ms{2}k{\setminus}n\ms{4}{\equiv}\ms{4}k{\setminus}x\ms{2}{\wedge}\ms{2}k{\setminus}y\rangle$\-\pop\\
	\textbf{Bound function}: $x{+}y$\-$ ~~\}$\pop\\
	\push$\MPsf{do}\ms{3}$\=\+$y\ms{1}{<}\ms{1}x\ms{2}\MPsf{\rightarrow}\ms{2}x\ms{1}:=\ms{1}x{-}y$\push\-\\
	$\MPsf{\Box}$	\>\+\pop$x\ms{1}{<}\ms{1}y\ms{2}\MPsf{\rightarrow}\ms{2}y\ms{1}:=\ms{1}y{-}x$\-\\
	$\MPsf{od}$\pop\\
	\push$\{~~$\=\+$0{<}x\ms{4}{\wedge}\ms{4}0{<}y\ms{4}{\wedge}\ms{4}x\ms{1}{=}\ms{1}y\ms{4}{\wedge}\ms{4}\langle\forall{}k{:}{:}\ms{3}k{\setminus}m\ms{2}{\wedge}\ms{2}k{\setminus}n\ms{4}{\equiv}\ms{4}k{\setminus}x\ms{2}{\wedge}\ms{2}k{\setminus}y\rangle$\-$ ~~\}$\pop
\end{mpdisplay}
The algorithm that we have constructed is  Euclid's algorithm for computing the greatest
common divisor of two positive natural numbers,  the oldest nontrivial algorithm that has survived 
to the present day! (Please note that our formulation of the algorithm differs from Euclid's original version and 
from most versions found in number-theory books. While they use the property $[\ms{3}m\ms{0}\raisebox{0.8mm}{\ensuremath{\bigtriangledown}} n\ms{3}{=}\ms{3}n\protect\raisebox{0.8mm}{$\bigtriangledown$}(m\ms{1}\mathsf{mod}\ms{1}n)\ms{3}]$, 
we use (\ref{prop-inv1}), i.e.,  $[\ms{3}m\ms{0}\raisebox{0.8mm}{\ensuremath{\bigtriangledown}} n\ms{3}{=}\ms{3}(m{-}n)\protect\raisebox{0.8mm}{$\bigtriangledown$}n\ms{3}]$. 
For an encyclopedic account of Euclid's algorithm, we recommend \cite[p. 334]{DEK-TAOCP2}.)

\subsection{Greatest Common Divisor}\label{gcd}
In section \ref{Division-Infimum}, we described the problem we were tackling as establishing that the infimum
of two natural numbers under the division ordering always exists;  it was only at the end of the section
that we announced that the algorithm we had derived is an algorithm for determining the greatest
common divisor.  This was done deliberately in order to avoid the confusion that can ---and does--- occur
when using the words ``greatest common divisor''.   In this section, we clarify the issue in some detail.

Confusion and ambiguity occur when a set can be ordered in two different ways.    The
natural numbers can be ordered by the usual size    ordering (denoted by the symbol $\leq$), but they can also be
ordered by the division relation.  When the ordering is not made explicit (for instance, when referring to
the ``least'' or ``greatest'' of a set of numbers), we might normally understand the size 
ordering, but the division ordering might be meant, depending on the context.  

In words, the infimum of two values in a partial ordering ---if it exists--- is the largest   value (with
respect to the ordering)  that is at most both values (with respect to the ordering).  The terminology
``greatest lower bound'' is often used instead of ``infimum''.  Of course, ``greatest'' here is with respect to
the partial ordering in question.  Thus,  the infimum (or greatest lower bound)  of two numbers with
respect to the division ordering ---if it exists--- is  the largest number with respect to the division
ordering  that divides both of the numbers.  Since, \emph{for strictly positive numbers},    ``largest with respect
to the division ordering''    implies  ``largest with respect to the size   ordering'' (equally, the division
relation, restricted to strictly positive numbers,  is a subset of the $\leq$ relation),  the  ``largest number with
respect to the \emph{division} ordering  that divides both of the numbers'' is the same, \emph{for strictly positive
numbers},   as the ``largest number with respect to the \emph{size}  ordering  that divides both of the numbers''.
 Both these expressions  may thus  be abbreviated  to the ``greatest common divisor'' of the numbers,
with no problems caused by the ambiguity in the meaning of ``greatest'' ---  \emph{when the numbers are 
strictly positive}.   Ambiguity  does occur, however,  when the number $0$ is included, because $0$ is the
\emph{largest} number with respect to the division ordering, but the \emph{smallest} number with respect to the size
ordering.     If ``greatest'' is taken to mean with respect to the division ordering on numbers, the greatest
common divisor of $0$ and $0$ is simply $0$. If, however, ``greatest'' is taken to
mean with respect to the size ordering, there is no greatest common divisor of $0$ and $0$.   
This would mean that the gcd  operator is no longer idempotent, since $0\protect\raisebox{0.8mm}{$\bigtriangledown$}0$ is 
undefined,  and  it is  no longer  associative, since, for positive $m$,   $(m\protect\raisebox{0.8mm}{$\bigtriangledown$}0)\protect\raisebox{0.8mm}{$\bigtriangledown$}0$ is well-defined whilst
$m\protect\raisebox{0.8mm}{$\bigtriangledown$}(0\protect\raisebox{0.8mm}{$\bigtriangledown$}0)$ is not.

Concrete evidence of the confusion in the standard mathematics literature is easy to find.
We looked up the definition of greatest common divisor in three commonly used undergraduate mathematics texts,
and found three non-equivalent definitions.  
The first \cite[p. 30]{Hirst95} defines ``greatest'' to  mean with respect to the divides relation (as, in our view, it should be
defined);  the second \cite[p. 21, def.\ 2.2]{burton2007-nt} defines ``greatest'' to mean with respect to the $\leq$ relation
(and requires that at least one of the numbers be non-zero).       The third text \cite[p. 78]{Fraleigh98}
excludes zero altogether, defining  the greatest common divisor of strictly positive numbers  as the
generator of all linear combinations of the given numbers; the accompanying  explanation (in words)  of
the terminology replaces ``greatest'' by ``largest'' but does not clarify with respect to which ordering the
``largest'' is to be determined.

Now that we know that $\protect\raisebox{0.8mm}{$\bigtriangledown$}$ is the greatest common divisor, we could change the operator to $\mathit{gcd\/}$, i.e.,
replace  $m\protect\raisebox{0.8mm}{$\bigtriangledown$}n$ by  $m~\mathit{gcd\/}~n$. However, we  stick to the ``$\protect\raisebox{0.8mm}{$\bigtriangledown$}$'' notation  because it makes the formulae
shorter, and, so, easier to read.   We also use ``$\bigtriangleup$'' to denote the least common multiple operator.   To
remember which is which, just remember that infima (\emph{lower} bounds) are indicated by
\emph{downward}-pointing symbols (eg.\ ${\downarrow}$ for minimum, and ${\vee}$ for disjunction) and  suprema (\emph{upper}
bounds)  by \emph{upward}-pointing symbols. 

\section{Euclid's Algorithm as a Verification Interface}\label{Euclid's Algorithm as a Verification Interface}
In this section we show  how algorithms and the notion of invariance can be used to prove theorems.
In particular, we show that the exploitation of Euclid's algorithm makes proofs
related with the greatest common divisor simple and more systematic than the 
traditional ones. 

There is a clear pattern in all our calculations: everytime we need
to prove a new theorem involving \setms{0.2em}$\protect\raisebox{0.8mm}{$\bigtriangledown$}$, we construct an invariant that is
valid initially (with $x\ms{1}{,}\ms{1}y\ms{1}:=\ms{1}m\ms{1}{,}\ms{1}n$) and that corresponds to the theorem to be proved
upon termination (with $x\ms{1}{=}\ms{1}y\ms{1}{=}\ms{1}m\protect\raisebox{0.8mm}{$\bigtriangledown$}n$). Alternatively, we can construct an
invariant that is valid on termination (with $x\ms{1}{=}\ms{1}y\ms{1}{=}\ms{1}m\protect\raisebox{0.8mm}{$\bigtriangledown$}n$) and whose initial value corresponds
to the theorem to be proved. The invariant in section \ref{A geometrical property} is such an example.
Then, it remains to prove that the chosen invariant is valid after each iteration of the repeatable statement.

We start with a minor change in the invariant that allows us to prove some well-known properties. Then, we
explore how the shape  of the theorems to be proved determine the shape of
the invariant. We also show how to prove a geometrical property of $\protect\raisebox{0.8mm}{$\bigtriangledown$}$.

\subsection{Exploring the invariant}
The invariant that we use in section \ref{Constructing Euclid's Algorithm} rests on the validity of the theorem
\begin{displaymath}[\ms{3}k{\setminus}m\ms{2}{\wedge}\ms{2}k{\setminus}n\ms{5}{\equiv}\ms{5}k{\setminus}(m{-}n)\ms{2}{\wedge}\ms{2}k{\setminus}n\ms{3}]\mbox{\ \ \ .}\end{displaymath}But, as Van Gasteren observed in \cite[Chapter 11]{netty90}, we can use the more general
and equally valid theorem
\begin{displaymath}[\ms{3}k\ms{1}{\setminus}\ms{1}(c{\MPtimes}m)\ms{2}{\wedge}\ms{2}k\ms{1}{\setminus}\ms{1}(c{\MPtimes}n)\ms{5}{\equiv}\ms{5}k\ms{1}{\setminus}\ms{1}(c\ms{1}{\MPtimes}\ms{1}(m{-}n))\ms{2}{\wedge}\ms{2}k\ms{1}{\setminus}\ms{1}(c{\MPtimes}n)\ms{3}]\end{displaymath}to conclude that the following property is an invariant of Euclid's algorithm:\begin{displaymath}\langle\forall{}k,c{:}{:}\ms{3}k\ms{1}{\setminus}\ms{1}(c{\MPtimes}m)\ms{2}{\wedge}\ms{2}k\ms{1}{\setminus}\ms{1}(c{\MPtimes}n)\ms{5}{\equiv}\ms{5}k\ms{1}{\setminus}\ms{1}(c{\MPtimes}x)\ms{2}{\wedge}\ms{2}k\ms{1}{\setminus}\ms{1}(c{\MPtimes}y)\rangle\mbox{\ \ \ .}\end{displaymath}In particular, the property is true on termination of the algorithm, at which point 
 $x$ and $y$ both equal 
$m\protect\raisebox{0.8mm}{$\bigtriangledown$}n$.  That is, for all $m$ and $n$, such that $0\ms{1}{<}\ms{1}m$ and  $0\ms{1}{<}\ms{1}n$,  
\begin{equation}\label{theorem-gcd-mult}
[\ms{3}k\ms{1}{\setminus}\ms{1}(c{\MPtimes}m)\ms{2}{\wedge}\ms{2}k\ms{1}{\setminus}\ms{1}(c{\MPtimes}n)\ms{6}{\equiv}\ms{6}k\ms{1}{\setminus}\ms{1}(c\ms{1}{\MPtimes}\ms{1}(m\protect\raisebox{0.8mm}{$\bigtriangledown$}n))\ms{3}]\mbox{\ \ \  . }
\end{equation}In addition,  theorem (\ref{theorem-gcd-mult}) holds when $m\ms{1}{<}\ms{1}0$, since\begin{displaymath}[\ms{3}(-m)\protect\raisebox{0.8mm}{$\bigtriangledown$}n\ms{1}{=}\ms{1}m\protect\raisebox{0.8mm}{$\bigtriangledown$}n\ms{3}]\ms{5}{\wedge}\ms{5}[\ms{3}k\ms{1}{\setminus}\ms{1}(c{\MPtimes}(-m))\ms{4}{\equiv}\ms{4}k\ms{1}{\setminus}\ms{1}(c{\MPtimes}m)\ms{3}]\mbox{\ \ \ ,}\end{displaymath}and it holds when $m$ equals $0$, since $[\ms{3}k{\setminus}0\ms{3}]$. 
Hence, using the symmetry between $m$ and $n$ we conclude that (\ref{theorem-gcd-mult}) is
indeed valid for all integers $m$ and $n$. (In Van Gasteren's
presentation, this theorem only holds for all $(m,n)\ms{1}{\neq}\ms{1}(0,0)$.) 

Theorem (\ref{theorem-gcd-mult}) can be used to prove a number of properties of
the greatest common divisor. If, for instance, we replace $k$ by $m$, we have\begin{displaymath}[\ms{3}m\ms{1}{\setminus}\ms{1}(c{\MPtimes}n)\ms{4}{\equiv}\ms{4}m\ms{1}{\setminus}\ms{1}(c\ms{1}{\MPtimes}\ms{1}(m\protect\raisebox{0.8mm}{$\bigtriangledown$}n))\ms{3}]\mbox{\ \ \ ,}\end{displaymath}and, as a consequence, we also have\begin{equation}\label{corollary1}
[\ms{3}(m\ms{1}{\setminus}\ms{1}(c{\MPtimes}n)\ms{4}{\equiv}\ms{4}m{\setminus}c)\ms{3}{\Leftarrow}\ms{3}m\protect\raisebox{0.8mm}{$\bigtriangledown$}n\ms{1}{=}\ms{1}1\ms{3}]\mbox{\ \ \ .}
\end{equation}More commonly,  (\ref{corollary1}) is formulated as the weaker\begin{displaymath}[\ms{3}m{\setminus}c\ms{4}{\Leftarrow}\ms{4}m\protect\raisebox{0.8mm}{$\bigtriangledown$}n\ms{1}{=}\ms{1}1\ms{2}{\wedge}\ms{2}m{\setminus}(c{\MPtimes}n)\ms{3}]\mbox{\ \ \ ,}\end{displaymath}and is known as Euclid's Lemma. Another significant property is
\begin{equation}\label{corollary2a}
[\ms{3}k\ms{1}{\setminus}\ms{1}(c\ms{1}{\MPtimes}\ms{1}(m\protect\raisebox{0.8mm}{$\bigtriangledown$}n))\ms{5}{\equiv}\ms{5}k\ms{1}{\setminus}\ms{1}((c{\times}m)\protect\raisebox{0.8mm}{$\bigtriangledown$}(c{\times}n))\ms{3}]\mbox{\ \ \ , }
\end{equation}which can be proved as:
\begin{mpdisplay}{0.2em}{5.1mm}{2mm}{2}
	$k\ms{1}{\setminus}\ms{1}(c\ms{1}{\MPtimes}\ms{1}(m\protect\raisebox{0.8mm}{$\bigtriangledown$}n))$\push\-\\
	$=$	\>	\>$\{$	\>\+\+\+(\ref{theorem-gcd-mult})\-\-$~~~ \}$\pop\\
	$k\ms{1}{\setminus}\ms{1}(c{\MPtimes}m)\ms{3}{\wedge}\ms{3}k\ms{1}{\setminus}\ms{1}(c{\MPtimes}n)$\push\-\\
	$=$	\>	\>$\{$	\>\+\+\+(\ref{nabla0})\-\-$~~~ \}$\pop\\
	$k\ms{1}{\setminus}\ms{1}((c{\times}m)\protect\raisebox{0.8mm}{$\bigtriangledown$}(c{\times}n))~~.$
\end{mpdisplay}
From (\ref{corollary2a}) we conclude\begin{equation}\label{corollary2}
[\ms{3}(c{\times}m)\protect\raisebox{0.8mm}{$\bigtriangledown$}(c{\times}n)\ms{1}{=}\ms{1}c\ms{1}{\MPtimes}\ms{1}(m\protect\raisebox{0.8mm}{$\bigtriangledown$}n)\ms{3}]\mbox{\ \ \ .}
\end{equation}Property (\ref{corollary2}) states that multiplication by  a 
natural number distributes over $\protect\raisebox{0.8mm}{$\bigtriangledown$}$.  It is an important
property that can be used to simplify arguments where both multiplication
and the greatest common divisor are involved. An example is  Van
Gasteren's proof of the theorem
\begin{equation}\label{prop-nabla-coprime}
[\ms{3}(m{\times}p)\protect\raisebox{0.8mm}{$\bigtriangledown$}n\ms{1}{=}\ms{1}m\protect\raisebox{0.8mm}{$\bigtriangledown$}n\ms{2}{\Leftarrow}\ms{2}p\protect\raisebox{0.8mm}{$\bigtriangledown$}n\ms{1}{=}\ms{1}1\ms{3}]\mbox{\ \ \ ,}
\end{equation}which is as follows:

\begin{mpdisplay}{0.2em}{5.1mm}{2mm}{2}
	$m\protect\raisebox{0.8mm}{$\bigtriangledown$}n$\push\-\\
	$=$	\>	\>$\{$	\>\+\+\+$p\protect\raisebox{0.8mm}{$\bigtriangledown$}n\ms{1}{=}\ms{1}1$ and 1 is the unit of multiplication\-\-$~~~ \}$\pop\\
	$(m{\times}(p\protect\raisebox{0.8mm}{$\bigtriangledown$}n))\protect\raisebox{0.8mm}{$\bigtriangledown$}n$\push\-\\
	$=$	\>	\>$\{$	\>\+\+\+(\ref{corollary2})\-\-$~~~ \}$\pop\\
	$(m{\times}p)\ms{1}\protect\raisebox{0.8mm}{$\bigtriangledown$}\ms{1}(m{\times}n)\ms{1}\protect\raisebox{0.8mm}{$\bigtriangledown$}\ms{1}n$\push\-\\
	$=$	\>	\>$\{$	\>\+\+\+$(m{\times}n)\protect\raisebox{0.8mm}{$\bigtriangledown$}n\ms{1}{=}\ms{1}n$\-\-$~~~ \}$\pop\\
	$(m{\times}p)\protect\raisebox{0.8mm}{$\bigtriangledown$}n~~.$
\end{mpdisplay}

\subsection{$\bigtriangledown$ on the left side}\label{Another invariant}
In the previous sections, we  have derived a number of properties of the $\protect\raisebox{0.8mm}{$\bigtriangledown$}$ operator.  However, where the
divides relation is involved,  the operator always occurs on the right side of the relation.   (For examples,
see (\ref{nabla0}) and
(\ref{corollary2a}).)     Now we consider properties where the operator is on the left side of a divides relation. 
Our goal is to show that\begin{equation}\label{theorem-linearcomb1}
[\ms{3}(m\protect\raisebox{0.8mm}{$\bigtriangledown$}n)\ms{1}{\setminus}\ms{1}k\ms{4}{\equiv}\ms{4}\langle\exists{}a,b{:}{:}\ms{3}k\ms{3}{=}\ms{3}m{\MPtimes}a\ms{1}{\MPplus}\ms{1}n{\MPtimes}b\rangle\ms{3}]\mbox{\ \ \ ,}
\end{equation}where the range of $a$ and $b$ is the integers.

Of course, if (\ref{theorem-linearcomb1}) is indeed true, then it is also true when $k$ equals $m\protect\raisebox{0.8mm}{$\bigtriangledown$}n$.  That is, a
consequence of (\ref{theorem-linearcomb1}) is\begin{equation}\label{theorem-linearcombcons}
[\ms{3}\langle\exists{}a,b{:}{:}\ms{3}m\protect\raisebox{0.8mm}{$\bigtriangledown$}n\ms{3}{=}\ms{3}m{\MPtimes}a\ms{1}{\MPplus}\ms{1}n{\MPtimes}b\rangle\ms{3}]\mbox{\ \ \ .}
\end{equation}In words, $m\protect\raisebox{0.8mm}{$\bigtriangledown$}n$ is a linear combination of $m$ and $n$.  For example,\begin{displaymath}3\protect\raisebox{0.8mm}{$\bigtriangledown$}5\ms{4}{=}\ms{4}1\ms{4}{=}\ms{4}3{\times}2\ms{1}{-}\ms{1}5{\times}1\ms{4}{=}\ms{4}5{\times}2\ms{1}{-}\ms{1}3{\times}3~~.\end{displaymath}Vice-versa, if (\ref{theorem-linearcombcons}) is indeed true then (\ref{theorem-linearcomb1}) is a consequence. 
(The crucial fact is that multiplication distributes through addition.)  It thus suffices to prove 
(\ref{theorem-linearcombcons}).

We can establish (\ref{theorem-linearcombcons}) by constructing such a linear combination for given values 
of $m$ and $n$.

When  $n$ is $0$, we have\begin{displaymath}m\protect\raisebox{0.8mm}{$\bigtriangledown$}0\ms{4}{=}\ms{4}m\ms{4}{=}\ms{4}m{\times}1\ms{1}{+}\ms{1}0{\times}1~~.\end{displaymath}(The multiple of $0$ is arbitrarily chosen to be $1$.)

When both $m$ and $n$ are non-zero, we need to augment Euclid's algorithm with a computation of the
coefficients.  The most effective way to establish the property is to establish that  $x$ and $y$ are linear
combinations of $m$ and $n$ is an invariant of the algorithm; this is best expressed using matrix arithmetic.

In the algorithm below,  the assignments to $x$ and $y$ have been replaced by equivalent assignments to 
the vector $(x\ms{2}y)$. Also, an additional variable $\mathbf{C\/}$, whose value is a $2{\times}2$ matrix of integers has been
introduced into the program.  Specifically,   $\mathbf{I\/}$, $\mathbf{A\/}$ and $\mathbf{B\/}$ are $2{\times}2$ matrices; $\mathbf{I\/}$ is the identity matrix $\left({1\atop 0}\;{0 \atop 1}\right)$ ,  
$\mathbf{A\/}$ is the matrix $\left({1\atop {-}1}\;{0 \atop 1}\right)$ and $\mathbf{B\/}$ is the matrix $\left({1\atop 0}\;{{-}1 \atop 1}\right)$. 
 (The assignment $(x\ms{2}y):=(x\ms{2}y){\times}\mathbf{A\/}$  is equivalent to $x\ms{1}{,}\ms{1}y\ms{1}:=\ms{1}x{-}y\ms{1}{,}\ms{1}y$, as can be easily checked.)
\label{algorithm-euclids-matrices}
\begin{mpdisplay}{0.2em}{5.1mm}{2mm}{2}
	\push$\{~~$\=\+$0\ms{1}{<}\ms{1}m\ms{3}{\wedge}\ms{3}0\ms{1}{<}\ms{1}n$\-$ ~~\}$\pop\\
	$(x\ms{2}y)\ms{1}{,}\ms{1}\mathbf{C\/}\ms{3}:=\ms{3}(m\ms{2}n)\ms{1}{,}\ms{1}\mathbf{I\/}\ms{2};$\\
	\push$\{~~$\=\+\push\textbf{Invariant}:$~~~~$\=\+$(x\ms{2}y)\ms{5}{=}\ms{5}(m\ms{2}n)\ms{1}{\times}\ms{1}\mathbf{C\/}$\-\pop\-$ ~~\}$\pop\\
	\push$\MPsf{do}\ms{3}$\=\+$y\ms{1}{<}\ms{1}x\ms{3}\MPsf{\rightarrow}\ms{3}(x\ms{2}y)\ms{1}{,}\ms{1}\mathbf{C\/}\ms{4}:=\ms{4}(x\ms{2}y)\ms{1}{\times}\ms{1}\mathbf{A\/}\ms{2}{,}\ms{2}\mathbf{C\/}{\times}\mathbf{A\/}$\push\-\\
	$\MPsf{\Box}$	\>\+\pop$x\ms{1}{<}\ms{1}y\ms{3}\MPsf{\rightarrow}\ms{3}(x\ms{2}y)\ms{1}{,}\ms{1}\mathbf{C\/}\ms{4}:=\ms{4}(x\ms{2}y)\ms{1}{\times}\ms{1}\mathbf{B\/}\ms{2}{,}\ms{2}\mathbf{C\/}{\times}\mathbf{B\/}$\-\\
	$\MPsf{od}$\pop\\
	\push$\{~~$\=\+$(x\ms{2}y)\ms{5}{=}\ms{5}(m\protect\raisebox{0.8mm}{$\bigtriangledown$}n\ms{3}m\protect\raisebox{0.8mm}{$\bigtriangledown$}n)\ms{5}{=}\ms{5}(m\ms{2}n)\ms{1}{\times}\ms{1}\mathbf{C\/}$\-$ ~~\}$\pop
\end{mpdisplay}
The invariant shows only the relation between the vectors $(x\ms{2}y)$ and $(m\ms{2}n)$; in words, $(x\ms{2}y)$ is a multiple
of $(m\ms{2}n)$.  

It is straightforward to verify that the invariant is established by the initialising assignment, and
maintained by the loop body.  Crucial to the proof  that it is maintained by the loop body is that
multiplication (here of matrices) is associative.  Had we expressed the assignments to $\mathbf{C\/}$ in terms of its
four elements, verifying that the invariant is maintained by the loop body would have amounted to
giving in detail the proof  that matrix multiplication is associative.    
This is a pointless duplication of effort, avoiding which
fully justifies the excursion into matrix arithmetic.

(An exercise for the reader is to express the property that $m$ and $n$ are linear combinations of  $x$ and $y$. 
The solution involves observing that $\mathbf{A\/}$ and $\mathbf{B\/}$ are invertible.  This will be exploited in section 
\ref{Enumerating the Rationals}.)

\subsection{A geometrical property}\label{A geometrical property}
In this section, we prove that in a Cartesian coordinate system, 
$m\protect\raisebox{0.8mm}{$\bigtriangledown$}n$ can be interpreted as the number of points with integral 
coordinates on the straight line joining the points $(0,0)$ and $(m,n)$, 
excluding $(0,0)$. Formally, with dummies  $s$ and $t$ ranging over integers, we prove for all $m$ and $n$:
\begin{equation}\label{nabla-geom0}
\begin {array}{lll}
&\langle\Sigma{}s,t\ms{3}{:}\ms{3}m{\times}t\ms{1}{=}\ms{1}n{\times}s\ms{3}{\wedge}\ms{3}s\ms{1}{\leq}\ms{1}m\ms{3}{\wedge}\ms{3}t\ms{1}{\leq}\ms{1}n\ms{2}{\wedge}\ms{2}(0\ms{1}{<}\ms{1}s\ms{1}{\vee}\ms{1}0\ms{1}{<}\ms{1}t)\ms{3}{:}\ms{3}1\rangle&\\
=&&\\
&m\protect\raisebox{0.8mm}{$\bigtriangledown$}n\mbox{\ \ \ .}&\\
\end{array}
\end{equation}
We begin by observing that (\ref{nabla-geom0}) holds when $m\ms{1}{=}\ms{1}0$ or when $n\ms{1}{=}\ms{1}0$ (we leave the proof to the reader).
When $0\ms{1}{<}\ms{1}m$ and $0\ms{1}{<}\ms{1}n$, we can simplify the range of (\ref{nabla-geom0}).
First, we observe that\begin{displaymath}(0\ms{1}{<}\ms{1}s\ms{1}{\leq}\ms{1}m\ms{4}{\equiv}\ms{4}0\ms{1}{<}\ms{1}t\ms{1}{\leq}\ms{1}n)\ms{2}{\Leftarrow}\ms{2}m{\times}t\ms{1}{=}\ms{1}n{\times}s\mbox{\ \ \ ,}\end{displaymath}since

\begin{mpdisplay}{0.2em}{5.1mm}{2mm}{2}
	$0\ms{1}{<}\ms{1}t\ms{1}{\leq}\ms{1}n$\push\-\\
	$=$	\>	\>$\{$	\>\+\+\+$0\ms{1}{<}\ms{1}m$\-\-$~~~ \}$\pop\\
	$0\ms{1}{<}\ms{1}m{\times}t\ms{1}{\leq}\ms{1}m{\times}n$\push\-\\
	$=$	\>	\>$\{$	\>\+\+\+$m{\times}t\ms{1}{=}\ms{1}n{\times}s$\-\-$~~~ \}$\pop\\
	$0\ms{1}{<}\ms{1}n{\times}s\ms{1}{\leq}\ms{1}m{\times}n$\push\-\\
	$=$	\>	\>$\{$	\>\+\+\+$0\ms{1}{<}\ms{1}n$,  cancellation \-\-$~~~ \}$\pop\\
	$0\ms{1}{<}\ms{1}s\ms{1}{\leq}\ms{1}m~~.$
\end{mpdisplay}
As a result, (\ref{nabla-geom0}) can be written as
\begin{equation}\label{nabla-geom1}
[\ms{3}\langle\Sigma{}s,t\ms{3}{:}\ms{3}m{\times}t\ms{1}{=}\ms{1}n{\times}s\ms{3}{\wedge}\ms{3}0\ms{1}{<}\ms{1}t\ms{1}{\leq}\ms{1}n\ms{3}{:}\ms{3}1\rangle\ms{1}{=}\ms{1}m\protect\raisebox{0.8mm}{$\bigtriangledown$}n\ms{3}]\mbox{\ \ \ .}
\end{equation}In order to use Euclid's algorithm, we need to find an invariant that allows us to conclude (\ref{nabla-geom1}). 
If we use as invariant
\begin{equation}\label{nabla-geom-invariant}
\langle\Sigma{}s,t\ms{3}{:}\ms{3}x{\times}t\ms{1}{=}\ms{1}y{\times}s\ms{3}{\wedge}\ms{3}0\ms{1}{<}\ms{1}t\ms{1}{\leq}\ms{1}y\ms{3}{:}\ms{3}1\rangle\ms{1}{=}\ms{1}x\protect\raisebox{0.8mm}{$\bigtriangledown$}y\mbox{\ \ \ ,}
\end{equation}its initial value is the property that we want to prove:\begin{displaymath}\langle\Sigma{}s,t\ms{3}{:}\ms{3}m{\times}t\ms{1}{=}\ms{1}n{\times}s\ms{2}{\wedge}\ms{2}0\ms{1}{<}\ms{1}t\ms{1}{\leq}\ms{1}n\ms{3}{:}\ms{3}1\rangle\ms{1}{=}\ms{1}m\protect\raisebox{0.8mm}{$\bigtriangledown$}n\mbox{\ \ \ .}\end{displaymath}Its value upon termination is\begin{displaymath}\langle\Sigma{}s,t\ms{3}{:}\ms{3}(m\protect\raisebox{0.8mm}{$\bigtriangledown$}n){\times}t\ms{1}{=}\ms{1}(m\protect\raisebox{0.8mm}{$\bigtriangledown$}n){\times}s\ms{2}{\wedge}\ms{2}0\ms{1}{<}\ms{1}t\ms{1}{\leq}\ms{1}m\ms{0}\raisebox{0.8mm}{\ensuremath{\bigtriangledown}} n\ms{3}{:}\ms{3}1\rangle\ms{1}{=}\ms{1}(m\protect\raisebox{0.8mm}{$\bigtriangledown$}n)\protect\raisebox{0.8mm}{$\bigtriangledown$}(m\protect\raisebox{0.8mm}{$\bigtriangledown$}n)\mbox{\ \ \ ,}\end{displaymath}which is equivalent (by cancellation of multiplication and idempotence of $\protect\raisebox{0.8mm}{$\bigtriangledown$}$) to\begin{displaymath}\langle\Sigma{}s,t\ms{3}{:}\ms{3}t\ms{1}{=}\ms{1}s\ms{2}{\wedge}\ms{2}0\ms{1}{<}\ms{1}t\ms{1}{\leq}\ms{1}m\protect\raisebox{0.8mm}{$\bigtriangledown$}n\ms{3}{:}\ms{3}1\rangle\ms{1}{=}\ms{1}m\protect\raisebox{0.8mm}{$\bigtriangledown$}n\mbox{\ \ \ .}\end{displaymath}It is easy to see that the invariant reduces to \textsf{true} on termination (because the sum on the left equals $m\protect\raisebox{0.8mm}{$\bigtriangledown$}n$), 
making its initial value also \textsf{true}.

It is also easy to see that the righthand side of the invariant  is unnecessary as it is the same initially and
on termination. This motivates the generalisation of the concept ``invariant''.
``Invariants'' in the literature are always boolean-valued functions of the program
variables, but we see no reason why ``invariants'' shouldn't be of any type: for us, an \emph{invariant} of a loop is
simply a function of the program variables whose value is unchanged by execution of the loop 
body\footnote{Some caution is needed here because our more general use of 
the word ``invariant'' does not  completely coincide with its standard usage for boolean-valued functions.
The standard meaning of an invariant of a statement $S$ is a boolean-valued function  of the program
variables which, in the case that the function evaluates to \textsf{true},  remains \textsf{true} after execution of $S$.  Our
usage requires that, if the function evaluates to \textsf{false} before execution of $S$, it continues to evaluate to \textsf{false}
after executing $S$.}.  In
this case, the value is a natural number.  Therefore, we can simplify (\ref{nabla-geom-invariant}) and use as invariant
\begin{equation}\label{nabla-geom-invariant1}
\langle\Sigma{}s,t\ms{3}{:}\ms{3}x{\times}t\ms{1}{=}\ms{1}y{\times}s\ms{3}{\wedge}\ms{3}0\ms{1}{<}\ms{1}t\ms{1}{\leq}\ms{1}y\ms{3}{:}\ms{3}1\rangle\mbox{\ \ \ .}
\end{equation}Its value on termination is\begin{displaymath}\langle\Sigma{}s,t\ms{3}{:}\ms{3}(m\protect\raisebox{0.8mm}{$\bigtriangledown$}n){\times}t\ms{1}{=}\ms{1}(m\protect\raisebox{0.8mm}{$\bigtriangledown$}n){\times}s\ms{2}{\wedge}\ms{2}0\ms{1}{<}\ms{1}t\ms{1}{\leq}\ms{1}m\ms{0}\raisebox{0.8mm}{\ensuremath{\bigtriangledown}} n\ms{3}{:}\ms{3}1\rangle\mbox{\ \ \ ,}\end{displaymath}which is equivalent to\begin{displaymath}\langle\Sigma{}s,t\ms{3}{:}\ms{3}t\ms{1}{=}\ms{1}s\ms{2}{\wedge}\ms{2}0\ms{1}{<}\ms{1}t\ms{1}{\leq}\ms{1}m\protect\raisebox{0.8mm}{$\bigtriangledown$}n\ms{3}{:}\ms{3}1\rangle\mbox{\ \ \ .}\end{displaymath}As said above, this sum equals $m\protect\raisebox{0.8mm}{$\bigtriangledown$}n$.

Now, since the invariant (\ref{nabla-geom-invariant1}) equals the lefthand side of (\ref{nabla-geom1}) for the initial values
of $x$ and $y$, we only have to check if it remains constant after each iteration. This means that
we have to prove (for $y\ms{1}{<}\ms{1}x\ms{1}{\wedge}\ms{1}0\ms{1}{<}\ms{1}y$):
\begin{mpdisplay}{0.2em}{5.1mm}{2mm}{2}
	\push$~~~\ms{4}$\=\+$\langle\Sigma{}s,t\ms{3}{:}\ms{3}x{\times}t\ms{1}{=}\ms{1}y{\times}s\ms{2}{\wedge}\ms{2}0\ms{1}{<}\ms{1}t\ms{1}{\leq}\ms{1}y\ms{3}{:}\ms{3}1\rangle$\push\-\\
	$=$	\>\+\pop$\langle\Sigma{}s,t\ms{3}{:}\ms{3}(x{-}y){\times}t\ms{1}{=}\ms{1}y{\times}s\ms{2}{\wedge}\ms{2}0\ms{1}{<}\ms{1}t\ms{1}{\leq}\ms{1}y\ms{3}{:}\ms{3}1\rangle$\,\,\,,\-\pop
\end{mpdisplay}
which can be rewritten, for positive $x$ and $y$, as:
\begin{mpdisplay}{0.2em}{5.1mm}{2mm}{2}
	\push$~~~\ms{4}$\=\+$\langle\Sigma{}s,t\ms{3}{:}\ms{3}(x{+}y){\times}t\ms{1}{=}\ms{1}y{\times}s\ms{2}{\wedge}\ms{2}0\ms{1}{<}\ms{1}t\ms{1}{\leq}\ms{1}y\ms{3}{:}\ms{3}1\rangle$\push\-\\
	$=$	\>\+\pop$\langle\Sigma{}s,t\ms{3}{:}\ms{3}x{\times}t\ms{1}{=}\ms{1}y{\times}s\ms{2}{\wedge}\ms{2}0\ms{1}{<}\ms{1}t\ms{1}{\leq}\ms{1}y\ms{3}{:}\ms{3}1\rangle~~.$\-\pop
\end{mpdisplay}
The proof is as follows:

\begin{mpdisplay}{0.2em}{5.1mm}{2mm}{2}
	$\langle\Sigma{}s,t\ms{3}{:}\ms{3}(x{+}y){\times}t\ms{1}{=}\ms{1}y{\times}s\ms{2}{\wedge}\ms{2}0\ms{1}{<}\ms{1}t\ms{1}{\leq}\ms{1}y\ms{3}{:}\ms{3}1\rangle$\push\-\\
	$=$	\>	\>$\{$	\>\+\+\+distributivity and cancellation\-\-$~~~ \}$\pop\\
	$\langle\Sigma{}s,t\ms{3}{:}\ms{3}x{\times}t\ms{1}{=}\ms{1}y{\times}(s{-}t)\ms{2}{\wedge}\ms{2}0\ms{1}{<}\ms{1}t\ms{1}{\leq}\ms{1}y\ms{3}{:}\ms{3}1\rangle$\push\-\\
	$=$	\>	\>$\{$	\>\+\+\+range translation: $s\ms{1}:=\ms{1}s{+}t$\-\-$~~~ \}$\pop\\
	$\langle\Sigma{}s,t\ms{3}{:}\ms{3}x{\times}t\ms{1}{=}\ms{1}y{\times}s\ms{2}{\wedge}\ms{2}0\ms{1}{<}\ms{1}t\ms{1}{\leq}\ms{1}y\ms{3}{:}\ms{3}1\rangle~~.$
\end{mpdisplay} 
Note that the simplification done in (\ref{nabla-geom1}) allows us to apply
the range translation rule in the last step without having to relate the range
of variable $s$ with the possible values for variable $t.$ 

\section{Euclid's Algorithm as a Construction Interface}\label{Euclid's Algorithm as a Construction Interface}
In this section we show  how to use Euclid's algorithm to derive new theorems  related with the greatest
common divisor. We start by calculating reasonable sufficient conditions for a natural-valued function to distribute
over the greatest common divisor. We also derive an efficient algorithm for enumerating the positive rational numbers
in two different ways.

\subsection{Distributivity properties}\label{Distributivity properties}
In addition  to  multiplication by a natural number, there are other
functions that  distribute over \setms{0.2em}$\protect\raisebox{0.8mm}{$\bigtriangledown$}$. The goal of this subsection is to 
determine sufficient conditions for a  natural-valued
function $f$ to distribute  over $\protect\raisebox{0.8mm}{$\bigtriangledown$}$, i.e., for  the following property to
hold:
\begin{equation}\label{nabla-dist0}
[\ms{3}f{.}(m\protect\raisebox{0.8mm}{$\bigtriangledown$}n)\ms{2}{=}\ms{2}f{.}m\ms{1}\protect\raisebox{0.8mm}{$\bigtriangledown$}\ms{1}f{.}n\ms{3}]\mbox{\ \ \ .}
\end{equation}For simplicity's sake, we restrict all variables to natural numbers. This implies that the domain of $f$
is also restricted to the natural numbers. 

We explore  (\ref{nabla-dist0}) by identifying  invariants of Euclid's algorithm involving
the function $f$. 
To determine an appropriate loop invariant, we take the right-hand side
of (\ref{nabla-dist0}) and calculate:
\begin{mpdisplay}{0.2em}{5.1mm}{2mm}{2}
	$f{.}m\ms{1}\protect\raisebox{0.8mm}{$\bigtriangledown$}\ms{1}f{.}n$\push\-\\
	$=$	\>	\>$\{$	\>\+\+\+the initial values of $x$ and $y$ are $m$ and $n$, respectively\-\-$~~~ \}$\pop\\
	$f{.}x\ms{1}\protect\raisebox{0.8mm}{$\bigtriangledown$}\ms{1}f{.}y$\push\-\\
	$=$	\>	\>$\{$	\>\+\+\+suppose that $f{.}x\ms{1}\protect\raisebox{0.8mm}{$\bigtriangledown$}\ms{1}f{.}y$ is invariant;\\
	on termination: $x\ms{1}{=}\ms{1}m\protect\raisebox{0.8mm}{$\bigtriangledown$}n\ms{2}{\wedge}\ms{2}y\ms{1}{=}\ms{1}m\protect\raisebox{0.8mm}{$\bigtriangledown$}n$\-\-$~~~ \}$\pop\\
	$f{.}(m\protect\raisebox{0.8mm}{$\bigtriangledown$}n)\ms{1}\protect\raisebox{0.8mm}{$\bigtriangledown$}\ms{1}f{.}(m\protect\raisebox{0.8mm}{$\bigtriangledown$}n)$\push\-\\
	$=$	\>	\>$\{$	\>\+\+\+$\protect\raisebox{0.8mm}{$\bigtriangledown$}$ is idempotent\-\-$~~~ \}$\pop\\
	$f{.}(m\protect\raisebox{0.8mm}{$\bigtriangledown$}n)~~.$
\end{mpdisplay}
Property (\ref{nabla-dist0}) is thus established under the assumption that $f{.}x\protect\raisebox{0.8mm}{$\bigtriangledown$}f{.}y$ is an invariant of
the loop body. (Please note that this invariant is of the more general form introduced in section \ref{A geometrical property}.)

The next step is to determine what condition  on $f$ guarantees that $f{.}x\protect\raisebox{0.8mm}{$\bigtriangledown$}f{.}y$ is indeed  
invariant. Noting the symmetry in the loop body between  $x$ and $y$, the condition is
easily calculated to be  \begin{displaymath}[\ms{3}f{.}(x{-}y)\ms{1}\protect\raisebox{0.8mm}{$\bigtriangledown$}\ms{1}f{.}y\ms{2}{=}\ms{2}f{.}x\ms{1}\protect\raisebox{0.8mm}{$\bigtriangledown$}\ms{1}f{.}y\ms{3}{\Leftarrow}\ms{3}0\ms{1}{<}\ms{1}y\ms{1}{<}\ms{1}x\ms{3}]\mbox{\ \ \ .}\end{displaymath}Equivalently, by the rule of range translation ($x\ms{1}:=\ms{1}x{+}y$), the condition can be written as\begin{equation}\label{nabla-dist-prop1}
[\ms{3}f{.}x\ms{1}\protect\raisebox{0.8mm}{$\bigtriangledown$}\ms{1}f{.}y\ms{2}{=}\ms{2}f{.}(x{+}y)\ms{1}\protect\raisebox{0.8mm}{$\bigtriangledown$}\ms{1}f{.}y\ms{3}{\Leftarrow}\ms{3}0\ms{1}{<}\ms{1}x\ms{2}{\wedge}\ms{2}0\ms{1}{<}\ms{1}y\ms{3}]\mbox{\ \ \ .}
\end{equation}Formally, this means that\begin{displaymath}\mbox{``\,}f\mbox{ distributes over }\protect\raisebox{0.8mm}{$\bigtriangledown$}\mbox{\,''\ms{3}}{\Leftarrow}\ms{3}\mbox{(\ref{nabla-dist-prop1})}\mbox{\ \ \ .}\end{displaymath}Incidentally, the converse of this property is also valid:\begin{displaymath}\mbox{(\ref{nabla-dist-prop1})\ms{2}}{\Leftarrow}\ms{2}\mbox{``\,}f\mbox{ distributes over }\protect\raisebox{0.8mm}{$\bigtriangledown$}\mbox{\,''}\mbox{\ \ \ .}\end{displaymath}The simple calculation proceeds as follows:
\begin{mpdisplay}{0.2em}{5.1mm}{2mm}{2}
	$f{.}(x{+}y)\ms{1}\protect\raisebox{0.8mm}{$\bigtriangledown$}\ms{1}f{.}y$\push\-\\
	$=$	\>	\>$\{$	\>\+\+\+$f$ distributes over $\protect\raisebox{0.8mm}{$\bigtriangledown$}$\-\-$~~~ \}$\pop\\
	$f{.}((x{+}y)\ms{0}\raisebox{0.8mm}{\ensuremath{\bigtriangledown}} y)$\push\-\\
	$=$	\>	\>$\{$	\>\+\+\+(\ref{nabla-lin-comb})\-\-$~~~ \}$\pop\\
	$f{.}(x\protect\raisebox{0.8mm}{$\bigtriangledown$}y)$\push\-\\
	$=$	\>	\>$\{$	\>\+\+\+$f$ distributes over $\protect\raisebox{0.8mm}{$\bigtriangledown$}$\-\-$~~~ \}$\pop\\
	$f{.}x\ms{1}\protect\raisebox{0.8mm}{$\bigtriangledown$}\ms{1}f{.}y~~.$
\end{mpdisplay}
By mutual implication we conclude that\begin{displaymath}\mbox{``\,}f\mbox{ distributes over }\protect\raisebox{0.8mm}{$\bigtriangledown$}\mbox{\,''\ms{5}}{\equiv}\ms{5}\mbox{(\ref{nabla-dist-prop1})}\mbox{\ \ \ .}\end{displaymath}We have now reached a point where we can determine if a function
distributes over $\protect\raisebox{0.8mm}{$\bigtriangledown$}$. However,  since (\ref{nabla-dist-prop1}) still has two occurrences of $\protect\raisebox{0.8mm}{$\bigtriangledown$}$, 
we want to refine it into simpler properties. Towards that
end we turn our attention to the condition
\begin{displaymath}f{.}x\ms{1}\protect\raisebox{0.8mm}{$\bigtriangledown$}\ms{1}f{.}y\ms{2}{=}\ms{2}f{.}(x{+}y)\ms{1}\protect\raisebox{0.8mm}{$\bigtriangledown$}\ms{1}f{.}y\mbox{\ \ \ ,}\end{displaymath}and we  explore simple ways of guaranteeing that it is everywhere true.
For instance,  it is immediately obvious that   any function that 
distributes over addition distributes over $\protect\raisebox{0.8mm}{$\bigtriangledown$}$.   (Note that 
multiplication by a natural number is such a function.) The proof is very simple:
\begin{mpdisplay}{0.2em}{5.1mm}{2mm}{2}
	$f{.}(x{+}y)\ms{1}\protect\raisebox{0.8mm}{$\bigtriangledown$}\ms{1}f{.}y$\push\-\\
	$=$	\>	\>$\{$	\>\+\+\+$f$ distributes over addition\-\-$~~~ \}$\pop\\
	$(f{.}x{+}f{.}y)\ms{1}\protect\raisebox{0.8mm}{$\bigtriangledown$}\ms{1}f{.}y$\push\-\\
	$=$	\>	\>$\{$	\>\+\+\+(\ref{nabla-lin-comb})\-\-$~~~ \}$\pop\\
	$f{.}x\ms{1}\protect\raisebox{0.8mm}{$\bigtriangledown$}\ms{1}f{.}y~~.$
\end{mpdisplay}
In view of properties (\ref{nabla-lin-comb}) and (\ref{prop-nabla-coprime}), 
we formulate the following lemma, which is a more general requirement:

\begin{Lemma}\label{nabla-dist-prop2}{\rm \ \ \ All functions $f$  that satisfy\begin{displaymath}\langle\forall{}x,y{:}{:}\ms{5}\langle\exists{}a,b\ms{3}{:}\ms{3}a\protect\raisebox{0.8mm}{$\bigtriangledown$}f{.}y\ms{2}{=}\ms{2}1\ms{3}{:}\ms{3}f{.}(x{+}y)\ms{3}{=}\ms{3}a\ms{1}{\times}\ms{1}f{.}x\ms{2}{+}\ms{2}b\ms{1}{\times}\ms{1}f{.}y\rangle\rangle\end{displaymath}distribute over $\protect\raisebox{0.8mm}{$\bigtriangledown$}$.
}
\end{Lemma}
{\bf Proof}~~~\begin{mpdisplay}{0.2em}{5.1mm}{2mm}{2}
	$f{.}(x{+}y)\ms{1}\protect\raisebox{0.8mm}{$\bigtriangledown$}\ms{1}f{.}y$\push\-\\
	$=$	\>	\>$\{$	\>\+\+\+$f{.}(x{+}y)\ms{3}{=}\ms{3}a\ms{1}{\times}\ms{1}f{.}x\ms{2}{+}\ms{2}b\ms{1}{\times}\ms{1}f{.}y$\-\-$~~~ \}$\pop\\
	$(a\ms{1}{\times}\ms{1}f{.}x\ms{2}{+}\ms{2}b\ms{1}{\times}\ms{1}f{.}y)\ms{1}\protect\raisebox{0.8mm}{$\bigtriangledown$}\ms{1}f{.}y$\push\-\\
	$=$	\>	\>$\{$	\>\+\+\+(\ref{nabla-lin-comb})\-\-$~~~ \}$\pop\\
	$(a\ms{1}{\times}\ms{1}f{.}x)\ms{1}\protect\raisebox{0.8mm}{$\bigtriangledown$}\ms{1}f{.}y$\push\-\\
	$=$	\>	\>$\{$	\>\+\+\+$a\ms{1}\protect\raisebox{0.8mm}{$\bigtriangledown$}\ms{1}f{.}y\ms{2}{=}\ms{2}1$ ~~~and~~~ (\ref{prop-nabla-coprime})\-\-$~~~ \}$\pop\\
	$f{.}x\ms{1}\protect\raisebox{0.8mm}{$\bigtriangledown$}\ms{1}f{.}y~~.$
\end{mpdisplay}
\nopagebreak\mnl$\Box$\ 

\noindent Note that since the discussion above is based on Euclid's algorithm, lemma \ref{nabla-dist-prop2} 
only applies to 
positive arguments. We now investigate the case where $m$ or $n$ is  $0$.   We have, for $m\ms{1}{=}\ms{1}0$ :

\begin{mpdisplay}{0.2em}{5.1mm}{2mm}{2}
	$f{.}(0\protect\raisebox{0.8mm}{$\bigtriangledown$}n)\ms{2}{=}\ms{2}f{.}0\ms{1}\protect\raisebox{0.8mm}{$\bigtriangledown$}\ms{1}f{.}n$\push\-\\
	$=$	\>	\>$\{$	\>\+\+\+$[\ms{3}0\protect\raisebox{0.8mm}{$\bigtriangledown$}m\ms{1}{=}\ms{1}m\ms{3}]$\-\-$~~~ \}$\pop\\
	$f{.}n\ms{2}{=}\ms{2}f{.}0\ms{1}\protect\raisebox{0.8mm}{$\bigtriangledown$}\ms{1}f{.}n$\push\-\\
	$=$	\>	\>$\{$	\>\+\+\+$[\ms{3}a{\setminus}b\ms{3}{\equiv}\ms{3}a\ms{1}{=}\ms{1}b\protect\raisebox{0.8mm}{$\bigtriangledown$}a\ms{3}]$\-\-$~~~ \}$\pop\\
	$f{.}n\ms{1}{\setminus}\ms{1}f{.}0$\push\-\\
	$\Leftarrow$	\>	\>$\{$	\>\+\+\+obvious possibilities that make the expression valid \\
	are  $f{.}0\ms{1}{=}\ms{1}0$,  $f{.}n\ms{1}{=}\ms{1}1$, or $f{.}n\ms{1}{=}\ms{1}f{.}0$; the first is the \\
	interesting case\-\-$~~~ \}$\pop\\
	$f{.}0\ms{1}{=}\ms{1}0~~.$
\end{mpdisplay}
Hence, using the symmetry between $m$ and $n$ we have, for $m\ms{1}{=}\ms{1}0$ or $n\ms{1}{=}\ms{1}0$:
\begin{equation}\label{prop-zero}
f{.}(m\protect\raisebox{0.8mm}{$\bigtriangledown$}n)\ms{2}{=}\ms{2}f{.}m\ms{1}\protect\raisebox{0.8mm}{$\bigtriangledown$}\ms{1}f{.}n\ms{5}{\Leftarrow}\ms{5}f{.}0\ms{1}{=}\ms{1}0\mbox{\ \ \ .}
\end{equation}The conclusion is that we can use (\ref{prop-zero}) and lemma \ref{nabla-dist-prop2}
to prove that a natural-valued function with domain $\MPNat$
distributes over $\protect\raisebox{0.8mm}{$\bigtriangledown$}$.
We  were unable to prove that the condition in lemma \ref{nabla-dist-prop2} is necessary for a function to distribute over
$\protect\raisebox{0.8mm}{$\bigtriangledown$}$, but we do not know any function distributing over $\protect\raisebox{0.8mm}{$\bigtriangledown$}$ that does not satisfy the condition.

\subsubsection*{Example 0: the Fibonacci function}
In \cite{EWD:EWD1077}, Edsger Dijkstra proves that the Fibonacci function
distributes over $\protect\raisebox{0.8mm}{$\bigtriangledown$}$ . He does not use lemma \ref{nabla-dist-prop2} explicitly, but 
he constructs the property
\begin{equation}\label{fib-prop1}
\mathsf{fib}{.}(x{+}y)\ms{3}{=}\ms{3}\mathsf{fib}{.}(y{-}1)\ms{1}{\times}\ms{1}\mathsf{fib}{.}x\ms{2}{+}\ms{2}\mathsf{fib}{.}(x{+}1)\ms{1}{\times}\ms{1}\mathsf{fib}{.}y\mbox{\ \ \ ,}
\end{equation}and then, using the lemma
\begin{displaymath}\mathsf{fib}{.}y\ms{1}\protect\raisebox{0.8mm}{$\bigtriangledown$}\ms{1}\mathsf{fib}{.}(y{-}1)\ms{2}{=}\ms{2}1\mbox{\ \ \ ,}\end{displaymath}he concludes the proof.   His calculation is the same as that in the proof of 
 lemma \ref{nabla-dist-prop2} but for 
particular values of  $a$ and  $b$ and with $f$ replaced by $\mathsf{fib}$.
Incidentally,  if we don't want to construct property (\ref{fib-prop1}) we can
easily verify it  using induction --- more details are given in 
\cite{Concrete2}.

An interesting application of this distributivity property is to prove that for any positive $k$, 
every $k$th number in the Fibonacci sequence is a multiple of the $k$th number
in the Fibonacci sequence. More formally, the goal is to prove
\begin{displaymath}\mathsf{fib}{.}(n{\times}k)\mbox{ is a multiple of }\mathsf{fib}{.}k\mbox{\ \ \ ,}\end{displaymath}for positive $k$ and natural $n.$ A  concise proof is:
\begin{mpdisplay}{0.2em}{5.1mm}{2mm}{2}
	$\mathsf{fib}{.}(n{\times}k)$ is a multiple of $\mathsf{fib}{.}k$\push\-\\
	$=$	\>	\>$\{$	\>\+\+\+definition\-\-$~~~ \}$\pop\\
	$\mathsf{fib}{.}k\ms{1}{\setminus}\ms{1}\mathsf{fib}{.}(n{\times}k)$\push\-\\
	$=$	\>	\>$\{$	\>\+\+\+$[\ms{3}a{\setminus}b\ms{3}{\equiv}\ms{3}a\protect\raisebox{0.8mm}{$\bigtriangledown$}b\ms{1}{=}\ms{1}a\ms{3}]$ , \\
	with $a\ms{1}{:=}\ms{1}\mathsf{fib}{.}k$ and $b\ms{1}{:=}\ms{1}\mathsf{fib}{.}(n{\times}k)$\-\-$~~~ \}$\pop\\
	$\mathsf{fib}{.}k\ms{1}\protect\raisebox{0.8mm}{$\bigtriangledown$}\ms{1}\mathsf{fib}{.}(n{\times}k)\ms{2}{=}\ms{2}\mathsf{fib}{.}k$\push\-\\
	$=$	\>	\>$\{$	\>\+\+\+$\mathsf{fib}$ distributes over $\protect\raisebox{0.8mm}{$\bigtriangledown$}$\-\-$~~~ \}$\pop\\
	$\mathsf{fib}{.}(k\protect\raisebox{0.8mm}{$\bigtriangledown$}(n{\times}k))\ms{1}{=}\ms{1}\mathsf{fib}{.}k$\push\-\\
	$=$	\>	\>$\{$	\>\+\+\+$k\protect\raisebox{0.8mm}{$\bigtriangledown$}(n{\times}k)\ms{1}{=}\ms{1}k$ and reflexivity\-\-$~~~ \}$\pop\\
	\textsf{true}$~~.$
\end{mpdisplay}

\subsubsection*{Example 1: the Mersenne function}
We now prove that, for all integers $k$ and $m$ such that $0\ms{1}{<}\ms{1}k^{m}$, 
the function $f$ defined as
\begin{displaymath}f{.}m\ms{3}{=}\ms{3}k^{m}{-}1\end{displaymath}distributes over $\protect\raisebox{0.8mm}{$\bigtriangledown$}$. 

First, we observe that $f{.}0\ms{1}{=}\ms{1}0$.  (Recall the discussion of (\ref{prop-zero}).)  
Next, we use lemma \ref{nabla-dist-prop2}.
This means that  we need to find integers $a$ and $b$, such that
\begin{displaymath}k^{m{+}n}{-}1\ms{3}{=}\ms{3}a{\times}(k^{m}{-}1)\ms{1}{+}\ms{1}b{\times}(k^{n}{-}1)\ms{5}{\wedge}\ms{5}a\protect\raisebox{0.8mm}{$\bigtriangledown$}(k^{n}{-}1)\ms{1}{=}\ms{1}1\mbox{\ \ \ .}\end{displaymath}The most obvious instantiations for $a$ are 1,  $k^{n}$ and $k^{n}{-}2$.   
(That two consecutive numbers are coprime  follows from (\ref{nabla-lin-comb}).) 
Choosing $a\ms{1}{=}\ms{1}1$, we calculate $b$:

\begin{mpdisplay}{0.2em}{5.1mm}{2mm}{2}
	$k^{m{+}n}{-}1\ms{3}{=}\ms{3}(k^{m}{-}1)\ms{1}{+}\ms{1}b{\times}(k^{n}{-}1)$\push\-\\
	$=$	\>	\>$\{$	\>\+\+\+arithmetic\-\-$~~~ \}$\pop\\
	$k^{m{+}n}{-}k^{m}\ms{2}{=}\ms{2}b{\times}(k^{n}{-}1)$\push\-\\
	$=$	\>	\>$\{$	\>\+\+\+multiplication distributes over addition\-\-$~~~ \}$\pop\\
	$k^{m}{\times}(k^{n}{-}1)\ms{2}{=}\ms{2}b{\times}(k^{n}{-}1)$\push\-\\
	$\Leftarrow$	\>	\>$\{$	\>\+\+\+Leibniz\-\-$~~~ \}$\pop\\
	$k^{m}\ms{1}{=}\ms{1}b~~.$
\end{mpdisplay}
We thus have\begin{displaymath}k^{m{+}n}{-}1\ms{2}{=}\ms{2}1{\times}(k^{m}{-}1)\ms{1}{+}\ms{1}k^{m}{\times}(k^{n}{-}1)\ms{3}{\wedge}\ms{3}1\protect\raisebox{0.8mm}{$\bigtriangledown$}(k^{n}{-}1)\ms{1}{=}\ms{1}1\mbox{\ \ \ ,}\end{displaymath}and we use lemma \ref{nabla-dist-prop2} to conclude that $f$ distributes over $\protect\raisebox{0.8mm}{$\bigtriangledown$}$:
\begin{displaymath}[\ms{3}(k^{m}{-}1)\ms{1}\protect\raisebox{0.8mm}{$\bigtriangledown$}\ms{1}(k^{n}{-}1)\ms{4}{=}\ms{4}k^{(m\ms{0}\tiny{\raisebox{0.5mm}{\ensuremath{\bigtriangledown}}} n)}{-}1\ms{3}]\mbox{\ \ \ .}\end{displaymath}In particular, the Mersenne function, which maps $m$ to  $2^{m}{-}1$, distributes over $\protect\raisebox{0.8mm}{$\bigtriangledown$}$:
\begin{equation}\label{mersenne-dist}
[\ms{3}(2^{m}{-}1)\ms{1}\protect\raisebox{0.8mm}{$\bigtriangledown$}\ms{1}(2^{n}{-}1)\ms{4}{=}\ms{4}2^{(m\ms{0}\tiny{\raisebox{0.5mm}{\ensuremath{\bigtriangledown}}} n)}{-}1\ms{3}]\mbox{\ \ \ .}
\end{equation}A corollary of (\ref{mersenne-dist}) is the property
\begin{displaymath}[\ms{3}(2^{m}{-}1)\protect\raisebox{0.8mm}{$\bigtriangledown$}(2^{n}{-}1)\ms{2}{=}\ms{2}1\ms{6}{\equiv}\ms{6}m\protect\raisebox{0.8mm}{$\bigtriangledown$}n\ms{1}{=}\ms{1}1\ms{3}]\mbox{\ \ \ .}\end{displaymath}In words, two numbers $2^{m}{-}1$ and $2^{n}{-}1$ are coprime is the same as
exponents $m$ and $n$ are coprime.

\renewcommand{\nabla}{\bigtriangledown}
\subsection{Enumerating the Rationals}\label{Enumerating the Rationals}
A standard theorem of mathematics is that the rationals are ``denumerable", i.e.\ they can be put in
one-to-one correspondence with the natural numbers.  Another way  of saying this is that it is possible to
enumerate the rationals so that each appears exactly once. 

Recently, there has been a spate of interest in the construction of  bijections between the natural numbers
and the (positive) rationals  (see \cite{Gibbons*2006:Enumerating,KnuthRupertSmithStong2003,Calkin-Wilf2000}
and \cite[pp. 94--97]{AZ2004}). Gibbons \emph{et al} 
\cite{Gibbons*2006:Enumerating}  describe as ``startling'' the observation that the rationals can be
efficiently enumerated\footnote{By an \emph{efficient enumeration} we mean a method of generating each
rational without duplication  with constant cost per rational 
in terms of arbitrary-precision simple arithmetic
operations.}  by  ``deforesting'' the so-called  ``Calkin-Wilf'' \cite{Calkin-Wilf2000} tree of
rationals.  However, they claim that it is ``not at all obvious'' how to ``deforest'' the Stern-Brocot tree of
rationals.   

In this section, we derive an  efficient  algorithm for enumerating the rationals according to both
orderings.   The algorithm is  based on a bijection between the rationals and invertible \setms{0.2em}$2{\times}2$ matrices.  The
key to the algorithm's derivation is the reformulation of Euclid's algorithm in terms of matrices (see
section \ref{algorithm-euclids-matrices}).  The enumeration is efficient  in the sense that it has the
same time and space complexity as  the algorithm credited to Moshe  Newman in 
\cite{KnuthRupertSmithStong2003}, albeit with a constant-fold increase in the number of variables and
number of arithmetic operations needed at each iteration.

Note that, in our view,  it is misleading to use the name  ``Calkin-Wilf tree of rationals'' because Stern 
\cite{stern1858:rationals}  had already documented essentially the same structural characterisation of the
rationals almost 150 years earlier than Calkin and Wilf.  For more explanation, see the appendix
 in which we review in some detail the relevant sections of Stern's paper.  Stern attributes the
structure to Eisenstein, so henceforth we refer to the ``Eisenstein-Stern'' tree of rationals where recent
publications (including our own \cite{jff*08:rationals}) would refer to the ``Calkin-Wilf tree of rationals''.  Section
 \ref{Brocot, the Watchmaker}  includes  background information. For a comprehensive account of properties
of the Stern-Brocot tree, including  further relationships with Euclid's algorithm, see 
\cite[pp. 116--118]{Concrete2}.

\subsubsection{Euclid's Algorithm}\label{Euclid's Algorithm}
A positive rational  in so-called ``lowest form'' is an ordered pair of positive, coprime  integers.  
Every rational $\frac{m}{n}$  has unique lowest-form representation $\frac{{}^{\smash{m}}\mskip-4mu/\mskip-3mu_{\smash{(m\nabla{}n)}}}{{}^{\smash{n}}\mskip-4mu/\mskip-3mu_{\smash{(m\nabla{}n)}}}$.   For example, $\frac{2}{3}$
is a rational in lowest form, whereas $\frac{4}{6}$ is the same rational, but not in lowest form.

Because computing the lowest-form representation involves computing greatest common divisors, 
it seems sensible to investigate Euclid's algorithm to see whether it gives insight into how to enumerate
the rationals.  Indeed it does.

Beginning with an arbitrary pair of positive integers $m$ and $n$, the algorithm presented in section
\ref{algorithm-euclids-matrices} calculates an invertible matrix $\mathbf{C\/}$ such that\begin{displaymath}(m\protect\raisebox{0.8mm}{$\bigtriangledown$}n\ms{3}m\protect\raisebox{0.8mm}{$\bigtriangledown$}n)\ms{5}{=}\ms{5}(m\ms{2}n)\ms{1}{\times}\ms{1}\mathbf{C\/}~~.\end{displaymath}It  follows that \begin{equation}\label{Cmn}
(1\ms{2}1)\ms{1}{\times}\ms{1}\mathbf{C\/}^{{-}1}\ms{5}{=}\ms{5}({}^{\smash{m}}\mskip-4mu/\mskip-3mu_{\smash{(m\nabla{}n)}}\ms{2}{}^{\smash{n}}\mskip-4mu/\mskip-3mu_{\smash{(m\nabla{}n)}})~~.
\end{equation}Because the  algorithm is deterministic,   positive  integers $m$ and $n$ uniquely define the matrix 
 $\mathbf{C\/}$.  That is, there is a function from pairs of positive integers to finite products of the matrices 
$\mathbf{A\/}$ and $\mathbf{B\/}$. Recall that $\mathbf{A\/}$ is the matrix $\left({1\atop {-}1}\;{0 \atop 1}\right)$ and $\mathbf{B\/}$ is the matrix $\left({1\atop 0}\;{{-}1 \atop 1}\right)$.  

Also, because the matrices $\mathbf{A\/}$ and $\mathbf{B\/}$ are constant and invertible,  $\mathbf{C\/}^{{-}1}$ is a  finite product  of the matrices
$\mathbf{A\/}^{{-}1}$ and $\mathbf{B\/}^{{-}1}$ and (\ref{Cmn})  uniquely defines   a rational ${\textstyle\frac{m}{n}}$.  We may therefore  conclude that there is a
bijection between the rationals and the finite products of the matrices $\mathbf{A\/}^{{-}1}$ and $\mathbf{B\/}^{{-}1}$ provided that we can
show that all such products are different.  

\begin{figure*}[h]
\includegraphics[width=\textwidth]{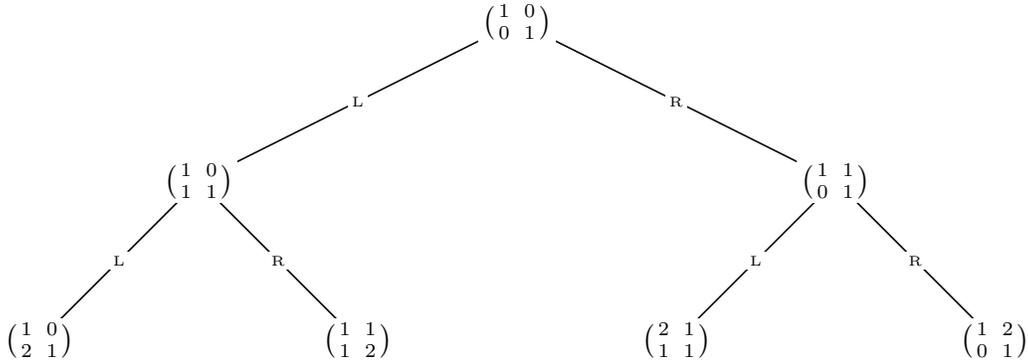} 
\caption{Tree of Products of $\mathbf{L\/}$  and $\mathbf{R\/}$}\label{matrixtree}
\end{figure*}

The finite products of  matrices $\mathbf{A\/}^{{-}1}$ and $\mathbf{B\/}^{{-}1}$ form a binary tree with root the identity matrix (the empty product).  
Renaming  $\mathbf{A\/}^{{-}1}$ as    $\mathbf{L\/}$ and $\mathbf{B\/}^{{-}1}$ as $\mathbf{R\/}$, the  tree can be displayed with ``L'' indicating a left branch and ``R'' indicating a right
branch.    Fig.\ \ref{matrixtree} displays the first few levels of the tree.

That all matrices in the tree are different is proved by showing that the tree is a binary search tree (as
formalised shortly).  The key element of the  proof\footnote{The proof is an
adaptation of the proof in \cite[p. 117]{Concrete2}  that the rationals in the Stern-Brocot tree are all different.   Our
use of determinants corresponds to their use of  ``the fundamental fact''  (4.31).
Note that  the definitions of  $\mathbf{L\/}$  and $\mathbf{R\/}$ are swapped around in  \cite{Concrete2}.)} is that the determinants of $\mathbf{A\/}$ and $\mathbf{B\/}$ are both equal to $1$
and, hence, the determinant of any finite product of $\mathbf{L\/}$s and $\mathbf{R\/}$s  is also $1$.  

Formally, we define the relation  ${\prec}$ on  matrices that are finite products of $\mathbf{L\/}$s and $\mathbf{R\/}$s  by   
\begin{displaymath}\begin{pmatrix}a&c\\b&d\end{pmatrix}\ms{2}{\prec}\ms{2}\left({a'\atop b'}\;{c' \atop d'}\right)\ms{8}{\equiv}\ms{8}\frac{a{+}c}{b{+}d}\ms{2}{<}\ms{2}\frac{a'{+}c'}{b'{+}d'}~~.\end{displaymath}  (Note  that the denominator in these fractions is strictly positive; this fact is easily proved by induction.)  
We prove that, for all such matrices  $\mathbf{X\/}$, $\mathbf{Y\/}$ and $\mathbf{Z\/}$,\begin{equation}\label{binsearch.ord}
\mathbf{X\/}{\times}\mathbf{L\/}{\times}\mathbf{Y\/}\ms{3}{\prec}\ms{3}\mathbf{X\/}\ms{3}{\prec}\ms{3}\mathbf{X\/}{\times}\mathbf{R\/}{\times}\mathbf{Z\/}~~.
\end{equation}It immediately  follows that there are no duplicates in the tree of matrices because the relation 
${\prec}$ is clearly transitive and a subset of the inequality relation.   
(Property (\ref{binsearch.ord}) formalises precisely what we mean by the tree of matrices forming a  binary
search tree:  the entries are properly ordered by the relation ${\prec}$, with matrices in the left branch being
``less than'' the root matrix which is ``less than'' matrices in the right branch.)  

In order to show that\begin{equation}\label{xly}
\mathbf{X\/}{\times}\mathbf{L\/}{\times}\mathbf{Y\/}\ms{2}{\prec}\ms{2}\mathbf{X\/}\mbox{\ \ \ ~~~,}
\end{equation}suppose $\mathbf{X\/}\ms{1}{=}\ms{1}\left({a\atop b}\;{c \atop d}\right)$ and $\mathbf{Y\/}\ms{1}{=}\ms{1}\left({a'\atop b'}\;{c' \atop d'}\right)$ . Then,  since $\mathbf{L\/}\ms{1}{=}\ms{1}\left({1\atop 1}\;{0 \atop 1}\right)$, 
(\ref{xly}) is easily calculated to be\begin{displaymath}\frac{(a{+}c){\times}a'\ms{1}{+}\ms{1}(c{\times}b')\ms{1}{+}\ms{1}(a{+}c){\times}c'\ms{1}{+}\ms{1}(c{\times}d')}{(b{+}d){\times}a'\ms{1}{+}\ms{1}(d{\times}b')\ms{1}{+}\ms{1}(b{+}d){\times}c'\ms{1}{+}\ms{1}(d{\times}d')}\ms{3}{<}\ms{3}\frac{a{+}c}{b{+}d}\mbox{\ \ \ .}\end{displaymath}That this is true is also a  simple, albeit longer,  calculation (which exploits the monotonicity properties of
multiplication and addition); as observed earlier, the  key property is that the determinant of $\mathbf{X\/}$ is $1$, i.e.\ 
$a{\times}d\ms{1}{-}\ms{1}b{\times}c\ms{2}{=}\ms{2}1$. The proof that $\mathbf{X\/}\ms{1}{\prec}\ms{1}\mathbf{X\/}{\times}\mathbf{R\/}{\times}\mathbf{Z\/}$ is similar.

Of course, we can also express Euclid's algorithm in terms of  transpose matrices.  Instead of writing
assignments to the vector $(x\ms{2}y)$, we can write assignments to its transpose
 $\left({x\atop y}\right)$.  Noting that  $\mathbf{A\/}$ and  $\mathbf{B\/}$ are each other's transposition, the assignment\begin{displaymath}(x\ms{2}y)\ms{1}{,}\ms{1}\mathbf{C\/}\ms{4}:=\ms{4}(x\ms{2}y)\ms{1}{\times}\ms{1}\mathbf{A\/}\ms{2}{,}\ms{2}\mathbf{C\/}{\times}\mathbf{A\/}\end{displaymath}in the body of Euclid's algorithm becomes \begin{displaymath}\left({x\atop y}\right)\ms{2}{,}\ms{2}\mathbf{C\/}\ms{6}:=\ms{6}\mathbf{B\/}\ms{1}{\times}\ms{1}\left({x\atop y}\right)\ms{3}{,}\ms{3}\mathbf{B\/}{\times}\mathbf{C\/}~~.\end{displaymath}Similarly, the assignment \begin{displaymath}(x\ms{2}y)\ms{1}{,}\ms{1}\mathbf{C\/}\ms{4}:=\ms{4}(x\ms{2}y)\ms{1}{\times}\ms{1}\mathbf{B\/}\ms{2}{,}\ms{2}\mathbf{C\/}{\times}\mathbf{B\/}\end{displaymath} becomes \begin{displaymath}\left({x\atop y}\right)\ms{2}{,}\ms{2}\mathbf{C\/}\ms{6}:=\ms{6}\mathbf{A\/}\ms{1}{\times}\ms{1}\left({x\atop y}\right)\ms{3}{,}\ms{3}\mathbf{A\/}{\times}\mathbf{C\/}~~.\end{displaymath}On termination, the matrix $\mathbf{C\/}$ computed by the revised algorithm will of course be different; the pair 
 $\left({{}^{\smash{m}}\mskip-4mu/\mskip-3mu_{\smash{(m\nabla{}n)}}\atop {}^{\smash{n}}\mskip-4mu/\mskip-3mu_{\smash{(m\nabla{}n)}}}\right)$ is recovered from it  by the identity\begin{displaymath}\mathbf{C\/}^{{-}1}\ms{1}{\times}\ms{1}\left({1\atop 1}\right)\ms{5}{=}\ms{5}\left({{}^{\smash{m}}\mskip-4mu/\mskip-3mu_{\smash{(m\nabla{}n)}}\atop {}^{\smash{n}}\mskip-4mu/\mskip-3mu_{\smash{(m\nabla{}n)}}}\right)~~.\end{displaymath}In this way, we get a second bijection between the rationals  and the finite products of the matrices $\mathbf{A\/}^{{-}1}$ 
and $\mathbf{B\/}^{{-}1}$.  This is the basis for our second method of enumerating the rationals.

In summary, we have:
\begin{Theorem}\label{rational-matrix bijection}{\rm \ \ \ Define the matrices  $\mathbf{L\/}$  and $\mathbf{R\/}$ by \begin{displaymath}\mathbf{L\/}\ms{4}{=}\ms{4}\left({1\atop 1}\;{0 \atop 1}\right)\textrm{~~~~and~~~~}\mathbf{R\/}\ms{4}{=}\ms{4}\left({1\atop 0}\;{1 \atop 1}\right)\mbox{\ \ \ .}\end{displaymath}Then the following algorithm computes a bijection between the (positive) rationals and the finite
products of $\mathbf{L\/}$  and $\mathbf{R\/}$. Specifically, the bijection is given by the function that maps the  rational $\frac{m}{n}$ to the
matrix $\mathbf{D\/}$ constructed by the algorithm together with the function from a finite product, $\mathbf{D\/}$,  of $\mathbf{L\/}$s  and $\mathbf{R\/}$s
to $(1\ms{2}1)\ms{1}{\times}\ms{1}\mathbf{D\/}$.    (The comments added to the algorithm supply the information needed to verify this
assertion.)

\begin{mpdisplay}{0.2em}{5.1mm}{2mm}{2}
	\push$\{~~$\=\+$0\ms{1}{<}\ms{1}m\ms{3}{\wedge}\ms{3}0\ms{1}{<}\ms{1}n$\-$ ~~\}$\pop\\
	$(x\ms{2}y)\ms{1}{,}\ms{1}\mathbf{D\/}\ms{3}:=\ms{3}(m\ms{2}n)\ms{1}{,}\ms{1}\mathbf{I\/}\ms{2};$\\
	\push$\{~~$\=\+\push\textbf{Invariant}:$~~~~$\=\+$(m\ms{2}n)\ms{5}{=}\ms{5}(x\ms{2}y)\ms{1}{\times}\ms{1}\mathbf{D\/}$\-\pop\-$ ~~\}$\pop\\
	\push$\MPsf{do}\ms{3}$\=\+$y\ms{1}{<}\ms{1}x\ms{3}\MPsf{\rightarrow}\ms{3}(x\ms{2}y)\ms{1}{,}\ms{1}\mathbf{D\/}\ms{4}:=\ms{4}(x\ms{2}y)\ms{1}{\times}\ms{1}\mathbf{L\/}^{{-}1}\ms{2}{,}\ms{2}\mathbf{L\/}{\times}\mathbf{D\/}$\push\-\\
	$\MPsf{\Box}$	\>\+\pop$x\ms{1}{<}\ms{1}y\ms{3}\MPsf{\rightarrow}\ms{3}(x\ms{2}y)\ms{1}{,}\ms{1}\mathbf{D\/}\ms{4}:=\ms{4}(x\ms{2}y)\ms{1}{\times}\ms{1}\mathbf{R\/}^{{-}1}\ms{2}{,}\ms{2}\mathbf{R\/}{\times}\mathbf{D\/}$\-\\
	$\MPsf{od}$\pop\\
	\push$\{~~$\=\+$(x\ms{2}y)\ms{4}{=}\ms{4}(m\protect\raisebox{0.8mm}{$\bigtriangledown$}n\ms{3}m\protect\raisebox{0.8mm}{$\bigtriangledown$}n)\ms{6}{\wedge}\ms{6}(\ms{1}{}^{\smash{m}}\mskip-4mu/\mskip-3mu_{\smash{(m\nabla{}n)}}\ms{2}{}^{\smash{n}}\mskip-4mu/\mskip-3mu_{\smash{(m\nabla{}n)}}\ms{1})\ms{4}{=}\ms{4}(1\ms{2}1)\ms{1}{\times}\ms{1}\mathbf{D\/}$\-$ ~~\}$\pop
\end{mpdisplay}

Similarly,  by applying  the rules of matrix transposition to all expressions in the above,  Euclid's
algorithm constructs a second bijection between the rationals and finite products of the matrices $\mathbf{L\/}$  and
$\mathbf{R\/}$.   Specifically, the bijection is given by the function that maps the rational $\frac{m}{n}$ to the matrix $\mathbf{D\/}$
constructed by the  revised  algorithm together with the function from  finite products, $\mathbf{D\/}$, of $\mathbf{L\/}$s  and $\mathbf{R\/}$s
to $\mathbf{D\/}\ms{1}{\times}\ms{1}\left({1\atop 1}\right)$.

}
\end{Theorem}
\nopagebreak$\Box$\

\subsubsection{Enumerating Products of $\mathbf{L\/}$  and $\mathbf{R\/}$}\label{Enumerating matrix products}

The problem of enumerating the rationals has been transformed to the problem of enumerating all finite
products of the matrices $\mathbf{L\/}$  and $\mathbf{R\/}$.    As observed earlier, the 
matrices are naturally visualised as a tree  ---recall fig.\  \ref{matrixtree}---
with  left branching corresponding to multiplying (on the right) by $\mathbf{L\/}$ and right branching to 
multiplying (on the right) by $\mathbf{R\/}$.  

\begin{figure}[h]
\includegraphics[width=\textwidth]{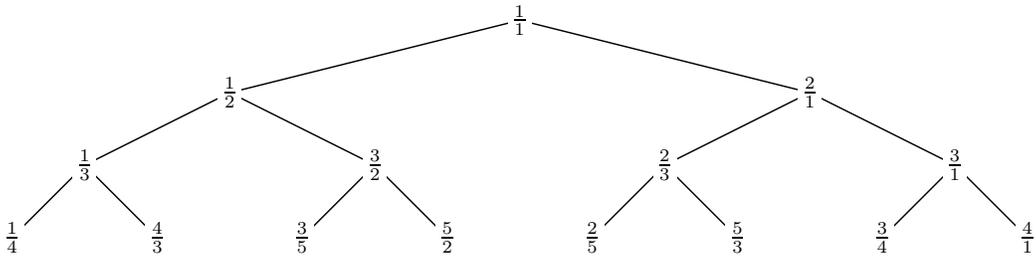} 
\caption{Eisenstein-Stern Tree of Rationals (aka Calkin-Wilf Tree)}\label{CalkinWilf}
\end{figure}

By premultiplying each matrix in the tree by $(1\ms{2}1)$, we get a tree of rationals.  (Premultiplying by $(1\ms{2}1)$ is
accomplished by adding the elements in each column.)  This tree is sometimes  called the Calkin-Wilf  tree 
\cite{Gibbons*2006:Enumerating,AZ2004,Calkin-Wilf2000}; we call it the \emph{Eisenstein-Stern tree} of rationals.  
(See the appendix for an explanation.)     The first four levels of the tree are
shown in  fig.  \ref{CalkinWilf}.  In this
figure, the vector $(x\ms{2}y)$ has been displayed as  $\frac{y}{x}$.    (Note the order  of $x$ and $y$.  This is  to
aid comparison with existing literature.)

\begin{figure}[h]
\includegraphics[width=\textwidth]{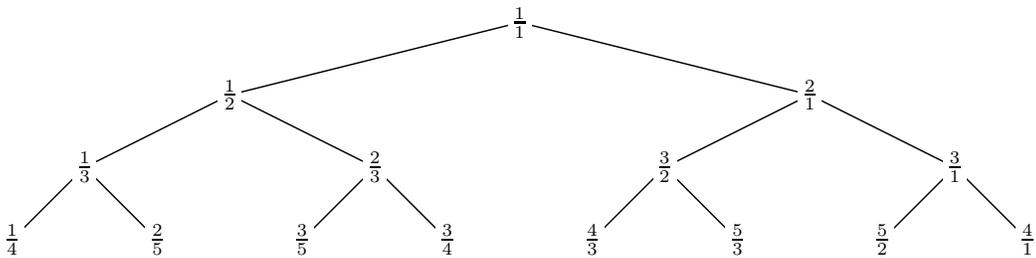} 
\caption{Stern-Brocot Tree of Rationals}\label{SternBrocot}
\end{figure}

By postmultiplying each matrix in the tree by $\left({1\atop 1}\right)$, we also get a tree of rationals.  (Postmultiplying by $\left({1\atop 1}\right)$ is
accomplished by adding the elements in each row.)  This tree is called the Stern-Brocot  tree 
\cite[pp. 116--118]{Concrete2}.     See fig.  \ref{SternBrocot}.  In this figure, the vector $\left({x\atop y}\right)$ 
has been displayed as $\frac{x}{y}$. 

Of course, if we can find an efficient  way of enumerating the matrices in fig.\ \ref{matrixtree}, we
immediately get an enumeration of the rationals as displayed in the Eisenstein-Stern tree 
and as displayed in the Stern-Brocot tree --- as each matrix is enumerated, simply premultiply by $(1\ms{2}1)$ or postmultiply by 
$\left({1\atop 1}\right)$.    Formally, the matrices are enumerated by enumerating all strings of Ls and Rs in lexicographic
order, beginning with the empty string;  each string is mapped to a matrix by the homomorphism  that
maps ``L'' to $\mathbf{L\/}$, ``R'' to $\mathbf{R\/}$, and  string concatenation to matrix product.  It is easy to enumerate all such
strings;  as we see shortly, converting strings to matrices   is also not  difficult,  for the simple reason that $\mathbf{L\/}$
and $\mathbf{R\/}$ are invertible.

The enumeration  proceeds level-by-level.  Beginning with the unit matrix (level $0$),
the matrices on each level are enumerated from left to right.  There are $2^{k}$ matrices on level $k$, the 
 first  of which is  $\mathbf{L\/}^{k}$.  The problem is to determine for a given matrix, which is the matrix ``adjacent'' to it.  
That is, given a matrix $\mathbf{D\/}$, which is a finite product of $\mathbf{L\/}$  and $\mathbf{R\/}$, and is different from $\mathbf{R\/}^{k}$ for all $k$, 
what is the matrix that  is to the immediate right of $\mathbf{D\/}$ in fig. \ref{matrixtree}?

Consider the lexicographic ordering on strings of Ls and Rs of  the same length.  
The string immediately following
a string $s$ (that is not the last) is found by identifying the rightmost L in $s$.  Supposing $s$ is the string $t$LR$^{j}$,  
where R$^{j}$ is a string of $j$ Rs, its successor is $t$RL$^{j}$.       

It's now easy to see how to transform the matrix identified by $s$ to its successor matrix.  Simply
postmultiply by $\mathbf{R\/}^{{-}j}\ms{1}{\times}\ms{1}\mathbf{L\/}^{{-}1}\ms{1}{\times}\ms{1}\mathbf{R\/}\ms{1}{\times}\ms{1}\mathbf{L\/}^{j}$.  This is because, for all $\mathbf{T\/}$ and $j$,\begin{displaymath}(\mathbf{T\/}\ms{1}{\times}\ms{1}\mathbf{L\/}\ms{1}{\times}\ms{1}\mathbf{R\/}^{j})\ms{1}{\times}\ms{1}(\mathbf{R\/}^{{-}j}\ms{1}{\times}\ms{1}\mathbf{L\/}^{{-}1}\ms{1}{\times}\ms{1}\mathbf{R\/}\ms{1}{\times}\ms{1}\mathbf{L\/}^{j})\ms{5}{=}\ms{5}\mathbf{T\/}\ms{1}{\times}\ms{1}\mathbf{R\/}\ms{1}{\times}\ms{1}\mathbf{L\/}^{j}~~.\end{displaymath}Also, it is easy to calculate $\mathbf{R\/}^{{-}j}\ms{1}{\times}\ms{1}\mathbf{L\/}^{{-}1}\ms{1}{\times}\ms{1}\mathbf{R\/}\ms{1}{\times}\ms{1}\mathbf{L\/}^{j}$.    Specifically,
\begin{displaymath}\mathbf{R\/}^{{-}j}\ms{1}{\times}\ms{1}\mathbf{L\/}^{{-}1}\ms{1}{\times}\ms{1}\mathbf{R\/}\ms{1}{\times}\ms{1}\mathbf{L\/}^{j}\ms{5}{=}\ms{5}\left({2{}j\ms{1}{+}\ms{1}1\atop {-}1}\;{1 \atop 0}\right)~~.\end{displaymath}(We omit the details.   Briefly, by induction,   $\mathbf{L\/}^{j}$ equals 
$\left({1\atop j}\;{0 \atop 1}\right)$.  Also, $\mathbf{R\/}$ is the transpose of $\mathbf{L\/}$.)  

The final task is to determine, given a matrix $\mathbf{D\/}$, which is a finite product of $\mathbf{L\/}$s  and $\mathbf{R\/}$s, and 
is different from $\mathbf{R\/}^{k}$ for all $k$,   the unique value $j$ such that $\mathbf{D\/}\ms{2}{=}\ms{2}\mathbf{T\/}\ms{1}{\times}\ms{1}\mathbf{L\/}\ms{1}{\times}\ms{1}\mathbf{R\/}^{j}$ for some $\mathbf{T\/}$.  This can be
determined by examining Euclid's algorithm once more.

The matrix form  of Euclid's algorithm discussed in theorem  \ref{rational-matrix bijection} 
computes a matrix $\mathbf{D\/}$ 
given a pair of   positive numbers $m$ and $n$; it maintains the invariant \begin{displaymath}(m\ms{2}n)\ms{5}{=}\ms{5}(x\ms{2}y)\ms{1}{\times}\ms{1}\mathbf{D\/}~~.\end{displaymath}$\mathbf{D\/}$ is initially the identity matrix and $x$ and $y$ are initialised to $m$ and $n$, respectively; 
immediately following the initialisation process,  
$\mathbf{D\/}$ is repeatedly premultiplied by $\mathbf{R\/}$ so long as $x$ is less than
$y$.  Simultaneously, $y$ is reduced by $x$.    
The  number of times that $\mathbf{D\/}$ is premultiplied by $\mathbf{R\/}$ is thus  the
greatest number $j$ such that $j{\times}m$  is less than $n$, which is  $\left\lfloor\frac{n{-}1}{m}\right\rfloor$.  Now
suppose the input values $m$ and $n$ are coprime.  Then, on termination of the algorithm, 
 $(1\ms{2}1)\ms{1}{\times}\ms{1}\mathbf{D\/}$  equals $(\ms{1}m\ms{2}n\ms{1})$.    That is, if \begin{displaymath}\mathbf{D\/}\ms{4}{=}\ms{4}\left({\mathbf{D\/}_{00}\atop \mathbf{D\/}_{10}}\;{\mathbf{D\/}_{01} \atop \mathbf{D\/}_{11}}\right)~~,\end{displaymath}then,\begin{displaymath}\left\lfloor\frac{n{-}1}{m}\right\rfloor\ms{4}{=}\ms{4}\left\lfloor\frac{\mathbf{D\/}_{01}\ms{1}{+}\ms{1}\mathbf{D\/}_{11}\ms{1}{-}\ms{1}1}{\mathbf{D\/}_{00}\ms{1}{+}\ms{1}\mathbf{D\/}_{10}}\right\rfloor~~.\end{displaymath}It remains  to decide how to keep track of the levels in the tree.  For this purpose,  it is not
necessary to maintain a counter.  It suffices to observe that
$\mathbf{D\/}$ is  a power of $\mathbf{R\/}$ exactly when the rationals in the Eisenstein-Stern, or Stern-Brocot, tree are integers, 
and this integer is the number of the next level in the tree (where the root is on level $0$).  
So, it is easy to test whether the last matrix on the current level has been reached.  
Equally, the first matrix
on the next level is easily calculated.  For reasons we discuss in the next section, 
we choose to test whether the rational
in the  Eisenstein-Stern tree is an integer; that is, we evaluate the boolean $\mathbf{D\/}_{00}\ms{1}{+}\ms{1}\mathbf{D\/}_{10}\ms{2}{=}\ms{2}1$. 
In this way, we get the following (non-terminating) program
which computes the successive values of $\mathbf{D\/}$.

\begin{mpdisplay}{0.2em}{5.1mm}{2mm}{2}
	$\mathbf{D\/}\ms{1}:=\ms{1}\mathbf{I\/}\ms{1};$\\
	\push$\MPsf{do}\ms{4}$\=\+$\mathbf{D\/}_{00}\ms{1}{+}\ms{1}\mathbf{D\/}_{10}\ms{2}{=}\ms{2}1\ms{4}\MPsf{\rightarrow}\ms{4}\mathbf{D\/}\ms{3}:=\ms{3}\left({1\atop \mathbf{D\/}_{01}{+}\mathbf{D\/}_{11}}\;{0 \atop 1}\right)$\push\-\\
	$\MPsf{\Box}$	\>\+\pop$\mathbf{D\/}_{00}\ms{1}{+}\ms{1}\mathbf{D\/}_{10}\ms{2}{\neq}\ms{2}1\ms{3}\MPsf{\rightarrow}\ms{3}j\ms{1}:=\ms{1}\left\lfloor{\textstyle\frac{\mathbf{D\/}_{01}\ms{1}{+}\ms{1}\mathbf{D\/}_{11}\ms{1}{-}\ms{1}1}{\mathbf{D\/}_{00}\ms{1}{+}\ms{1}\mathbf{D\/}_{10}}}\right\rfloor\ms{3};$\\
	\push$~~~$\=\+\push$~~~$\=\+\push$~~~$\=\+\push$~~~$\=\+\push$~~~$\=\+\push$~~~$\=\+\push$~~~$\=\+\push$~~~\ms{2}$\=\+$\raisebox{6mm}{}\mathbf{D\/}\ms{1}:=\ms{1}\mathbf{D\/}\ms{1}{\times}\ms{1}\left({2{}j\ms{1}{+}\ms{1}1\atop {-}1}\;{1 \atop 0}\right)$\-\pop\-\pop\-\pop\-\pop\-\pop\-\pop\-\pop\-\pop\-\\
	$\MPsf{od}$\pop
\end{mpdisplay}
A minor simplification of this algorithm is that the ``${-}\ms{1}1$'' in the assignment to  $j$ can be omitted.  This is
because $\left\lfloor\frac{n{-}1}{m}\right\rfloor$ and $\left\lfloor\frac{n}{m}\right\rfloor$ are equal when $m$ and $n$ are coprime and $m$ is different from $1$. 
We return to  this shortly. 

\subsubsection{The Enumerations}\label{The Enumerations}

As remarked earlier,  we
immediately get an enumeration of the rationals as displayed in the Eisenstein-Stern tree 
and as displayed in the Stern-Brocot tree --- as each matrix is enumerated, 
simply premultiply by $(1\ms{2}1)$ or postmultiply by $\left({1\atop 1}\right)$, respectively.  

In the case of enumerating the Eisenstein-Stern tree, several optimisations are possible.  First, it is immediate
from our derivation that the value assigned to the local variable $j$ is a function of $(1\ms{2}1)\ms{1}{\times}\ms{1}\mathbf{D\/}$.  In turn, the
matrix $\left({2{}j\ms{1}{+}\ms{1}1\atop {-}1}\;{1 \atop 0}\right)$ is also a function of $(1\ms{2}1)\ms{1}{\times}\ms{1}\mathbf{D\/}$.  Let us name
the function $J$, so that the assignment becomes \begin{displaymath}\mathbf{D\/}\ms{3}:=\ms{3}\mathbf{D\/}\ms{2}{\times}\ms{2}J{.}((1\ms{2}1)\ms{1}{\times}\ms{1}\mathbf{D\/})~~.\end{displaymath}Then, the Eisenstein-Stern enumeration iteratively evaluates \begin{displaymath}(1\ms{2}1)\ms{1}{\times}\ms{1}(\mathbf{D\/}\ms{2}{\times}\ms{2}J{.}((1\ms{2}1)\ms{1}{\times}\ms{1}\mathbf{D\/}))~~.\end{displaymath}Matrix multiplication is associative; so this is \begin{displaymath}((1\ms{2}1)\ms{1}{\times}\ms{1}\mathbf{D\/})\ms{2}{\times}\ms{2}J{.}((1\ms{2}1)\ms{1}{\times}\ms{1}\mathbf{D\/})~~,\end{displaymath}which is also a function of $(1\ms{2}1)\ms{1}{\times}\ms{1}\mathbf{D\/}$.    Moreover ---in anticipation of the current discussion---  we have been
careful to ensure that the test for a change in the level in the tree is also a function of $(1\ms{2}1)\ms{1}{\times}\ms{1}\mathbf{D\/}$.  
Combined together, this means that, in order to enumerate the rationals in Eisenstein-Stern
order, it is not necessary to compute $\mathbf{D\/}$ at each iteration, but only $(1\ms{2}1)\ms{1}{\times}\ms{1}\mathbf{D\/}$.    Naming the two
components of this vector $m$ and $n$, and simplifying the matrix multiplications, we get\footnote{Recall
that, to comply with existing literature, the enumerated rational is $\frac{n}{m}$ and not $\frac{m}{n}$.}
\begin{mpdisplay}{0.2em}{5.1mm}{2mm}{2}
	$m{,}n\ms{2}:=\ms{2}1{,}1\ms{2};$\\
	\push$\MPsf{do}\ms{4}$\=\+$m\ms{1}{=}\ms{1}1\ms{5}\MPsf{\rightarrow}\ms{5}m{,}n\ms{4}:=\ms{4}n{+}1\ms{1}{,}\ms{1}m$\push\-\\
	$\MPsf{\Box}$	\>\+\pop$m\ms{1}{\neq}\ms{1}1\ms{5}\MPsf{\rightarrow}\ms{5}m{,}n\ms{4}:=\ms{4}(2{}\left\lfloor{\displaystyle\frac{n\ms{1}{-}\ms{1}1}{m}}\right\rfloor\ms{1}{+}\ms{1}1)\ms{1}{\times}\ms{1}m\ms{1}{-}\ms{1}n\ms{3}{,}\ms{3}m$\-\\
	$\MPsf{od}$\pop
\end{mpdisplay}
At this point, a further simplification is also possible.  We remarked earlier that $\left\lfloor{\textstyle\frac{n\ms{1}{-}\ms{1}1}{m}}\right\rfloor$ equals $\left\lfloor{\textstyle\frac{n}{m}}\right\rfloor$ when
$m$ and $n$ are coprime and $m$ is different from $1$.  By good fortune, it is also the case that 
$(2{}\left\lfloor{\textstyle\frac{n}{m}}\right\rfloor\ms{1}{+}\ms{1}1)\ms{1}{\times}\ms{1}m\ms{1}{-}\ms{1}n$ simplifies to $n{+}1$ when $m$ is equal to $1$.    That is, the elimination of  ``${-}\ms{1}1$'' in the
evaluation of the floor function leads to  
the elimination of the entire case analysis!
This is the algorithm attributed to   Newman in  \cite{KnuthRupertSmithStong2003}.
\begin{mpdisplay}{0.2em}{5.1mm}{2mm}{2}
	$m{,}n\ms{2}:=\ms{2}1{,}1\ms{2};$\\
	\push$\MPsf{do}\ms{7}$\=\+$m{,}n\ms{4}:=\ms{4}(2{}\left\lfloor{\displaystyle\frac{n}{m}}\right\rfloor\ms{1}{+}\ms{1}1)\ms{1}{\times}\ms{1}m\ms{1}{-}\ms{1}n\ms{3}{,}\ms{3}m$\-\\
	$\MPsf{od}$\pop
\end{mpdisplay}

\subsubsection{Discussion}\label{Rationals Discussion}
Our construction of an algorithm for enumerating the rationals in Stern-Brocot order 
was motivated by reading two publications,  \cite[pp. 116--118]{Concrete2} and
\cite{Gibbons*2006:Enumerating}.  Gibbons, Lester and
Bird \cite{Gibbons*2006:Enumerating}  show how to enumerate the elements of the Eisenstein-Stern tree, 
but claim  that ``it is not at all obvious
how to do this for the Stern-Brocot tree".  Specifically, they say\footnote{Recall that they attribute the
tree to Calkin and Wilf rather than Eisenstein and Stern.}:
\begin{quote}
However, there is an even better compensation for the loss of the
ordering property in moving from the Stern-Brocot to the Calkin-Wilf
tree: it becomes possible to deforest the tree altogether, and
generate the rationals directly, maintaining no additional state
beyond the `current' rational. This startling observation is due to
Moshe Newman (Newman, 2003). In contrast, it is not at
all obvious how to do this for the Stern-Brocot tree; the best we can
do seems to be to deforest the tree as far as its levels, but this
still entails additional state of increasing size.\end{quote}
In this section, we have shown that it is possible to  enumerate the rationals in
Stern-Brocot order without incurring ``additional state of increasing size''.  More importantly, we have 
presented \emph{one} enumeration algorithm  with \emph{two} specialisations, one being the
``Calkin-Wilf'' enumeration they present, and the other being the Stern-Brocot enumeration that they
described as being ``not at all obvious''.  

The optimisation of  Eisenstein-Stern enumeration which leads to Newman's algorithm  is
 not possible for  Stern-Brocot enumeration.  Nevertheless, the complexity of
Stern-Brocot enumeration is the same as the complexity of 
Newman's algorithm, both in time and space.   The only disadvantage of 
Stern-Brocot enumeration is that four variables are needed in place of two; the advantage is the
(well-known) advantage of the Stern-Brocot tree over the Eisenstein-Stern tree --- 
the rationals on a given level are in ascending order.

Gibbons, Lester and Bird's goal seems to have been  to show how 
the functional programming language Haskell implements the various
constructions -- the construction of the tree structures and Newman's algorithm.  In doing so, they
repeat the existing mathematical presentations of the algorithms as
given in \cite{Concrete2, Calkin-Wilf2000, KnuthRupertSmithStong2003}.  The ingredients for an efficient
enumeration of the Stern-Brocot tree are all present in these publications, but the recipe is missing!

The fact that  expressing
the rationals in ``lowest form'' is  essential to the avoidance of  duplication in any
enumeration immediately suggests the relevance of Euclid's algorithm.  
The key  to our exposition is that 
 Euclid's algorithm can be expressed in terms of matrix multiplications, where
---significantly--- the underlying matrices are invertible.  Transposition and inversion of the
matrices capture  the symmetry properties in a precise, calculational framework.  As a result, the
bijection between the rationals and the tree elements is immediate and we do not need to give
separate, inductive proofs for both tree structures.  Also, the determination of the next element in
an enumeration of the tree elements  has been reduced to one unifying  construction.

\section{Conclusion}\label{Discussion}

In our view, much of mathematics is inherently algorithmic; it is also clear that, in the modern age,
algorithmic problem solving is  just as important, if not much more so, than in the 19th century.  
Somehow, however, mathematical education in the 20th century lost sight of its algorithmic
roots.  We hope to have exemplified in this paper how a fresh approach to introductory number theory that
focuses on the algorithmic content of the theory can combine practicality with  mathematical  elegance.
By continuing this endeavour we believe that the teaching of mathematics can be enriched and
given new vigour.


\bibliographystyle{elsarticle-num}
\bibliography{bibliogr,ewd,math,knuth,jff}
\part*{Appendix:  Historical Remarks}\label{Appendix:  Historical Remarks}

The primary novel result of our paper is the 
construction given in section \ref{Enumerating the Rationals} of an algorithm to enumerate the rationals in
Stern-Brocot order.  Apart from minor differences, 
this section of our paper  was submitted in April 2007 to the American 
Mathematical Monthly; it was rejected in
November 2007 on the grounds that it was not of sufficient interest to readers of the Monthly.  One (of two
referees) did, however, recommend publication.  The referee made the following general comment.
\begin{quote}  Each of the two trees of rationals---the Stern-Brocot tree and the Calkin-Wilf tree---has
some history.  Since this paper now gives the definitive link between these trees, I encourage the authors,
perhaps in their Discussion section, to also give the definitive histories of these trees, something in the
same spirit as the Remarks at the end of the Calkin and Wilf paper. \end{quote}
Since the publication of \cite{jff*08:rationals}, we have  succeeded in obtaining copies of the original papers and it is indeed interesting to briefly
review the papers.   But we  do not claim to provide 
``definitive histories of these trees''  --- that is a task for a historian of mathematics.

Section \ref{Stern's Paper} is about the paper \cite{stern1858:rationals} published in 1858 by Stern.  The
surprising fact that emerges from the review is that the so-called ``Calkin-Wilf'' tree of rationals, and
not just the ``Stern-Brocot'' tree,  is  studied in detail in his paper.  Moreover, of the two structures,  the
``Calkin-Wilf'' tree is  more readily recognised; the ``Stern-Brocot'' tree requires rather more
understanding to identify.  Brocot's paper \cite{brocot-original}, which we review in section \ref{Brocot, the
Watchmaker},  is interesting because it illustrates how 19th century mathematics was driven by
practical, algorithmic problems. (For additional historical remarks, see also \cite{hayes00:teeth-wheels}.)

\section{Stern's Paper}\label{Stern's Paper}

Earlier we have commented that the structure that has recently been referred to as the ``Calkin-Wilf'' 
tree  was documented by Stern \cite{stern1858:rationals} in 1858. In this section we review those sections of
Stern's paper that are relevant to our own.

\subsection{The Eisenstein Array}\label{The Eisenstein Array}

Stern's paper is a detailed study of what has now become known as the ``Eisenstein array'' of numbers
(see, for example, \cite[sequence A064881]{sloane-integers}). 
(Stern's paper cites two papers written by the more famous mathematician Gotthold Eisenstein; 
we have not read these papers.)  Given two natural
numbers $m$ and $n$, Stern describes a process (which he attributes to Eisenstein) 
of generating an infinite  sequence of rows of numbers.  
The \emph{zeroth}  row in the sequence (``nullte Entwickelungsreihe'')   is the given pair of numbers:\begin{displaymath}
m\ms{5}n\mbox{\ \ \ .}
\end{displaymath}
Subsequent rows are obtained by inserting between every pair of numbers the sum of the numbers.  Thus
the \emph{first} row is\begin{displaymath}
m\ms{6}\setms{0.2em}m{+}n\ms{5}n 
\end{displaymath}
and the \emph{second} row is\begin{displaymath}
m\ms{6}2{\times}m\ms{1}{+}\ms{1}n\ms{6}m{+}n\ms{6}m\ms{1}{+}\ms{1}2{\times}n\ms{6}n~~. 
\end{displaymath}
The process of constructing such rows is repeated indefinitely.  The sequence of numbers obtained by
concatenating the individual rows in order is what is now called the \emph{Eisenstein array} and denoted by
 $\mathit{Ei\/}(m{,}n)$ (see, for example, \cite[sequence A064881]{sloane-integers}) .  
Stern refers to each occurrence of a number  in rows other than the
zeroth row as either  a 
\emph{sum element} (``Summenglied'')  or a  \emph{source element} (``Stammglied'').    The sum elements are the
newly added numbers.  For example, in the first row the
number $m{+}n$ is a sum element; in the second row the number   $m{+}n$ is a  source element. 

\subsection{The Eisenstein-Stern Tree of Rationals}\label{The Eisenstein-Stern Tree of Rationals}

A central element of Stern's analysis of the Eisenstein array is the consideration of subsequences of
 numbers in individual rows.   He calls these \emph{groups}  (``Gruppen'') and he
records the properties of pairs of consecutive numbers (groups of size two --- ``zweigliedrige Gruppen'') 
and triples of  consecutive numbers (groups of size three --- ``dreigliedrige Gruppen'').  

\begin{figure}[h]
\begin{center}\includegraphics[width=0.6\textwidth]{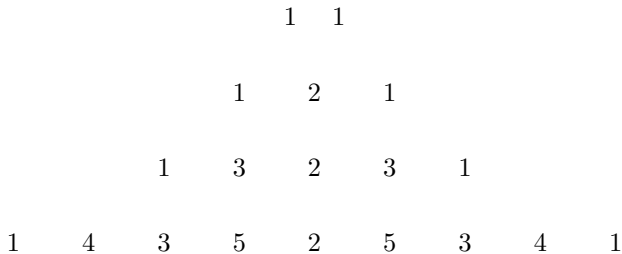}\end{center} 
\caption{First four rows of $\mathit{Ei\/}(1{,}1)$}\label{E11 rows}
\end{figure}

\begin{figure}[h]
\begin{center}\includegraphics[width=0.6\textwidth]{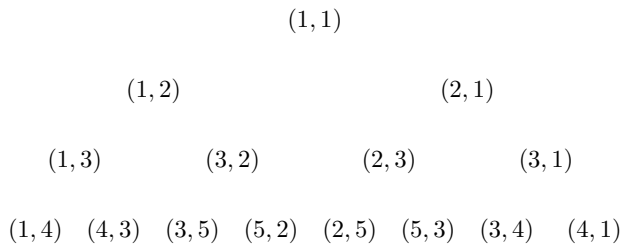}\end{center} 
\caption{Pairs of consecutive numbers in the first four rows of $\mathit{Ei\/}(1{,}1)$}\label{Pairs consecutive numbers}
\end{figure}

In sections 5 thru 8 of his paper, Stern studies $\mathit{Ei\/}(1{,}1)$,  the Eisenstein array that begins with the pair $(1,1)$.
  He  proves that all pairs of consecutive numbers in a given row  are coprime and every pair of coprime
numbers appears exactly once as such a pair of consecutive numbers.  He does not use the word ``tree'' 
---tree structures are most probably an invention of modern computing science--- and he does not refer
to ``rational numbers''  ---he refers instead to relatively prime numbers (``relatieve Primzahlen'')--- 
but there is no doubt that, apart from the  change in terminology,  he describes the tree of rationals that
in recent years has been referred to as the ``Calkin-Wilf'' tree of rationals.   It is for this reason that we
believe it is misleading to use the name ``Calkin-Wilf tree'' and prefer to use the name
``Eisenstein-Stern tree''. Fig. \ref{E11 rows} shows the first four rows of $\mathit{Ei\/}(1{,}1)$ and fig. \ref{Pairs consecutive numbers}
shows all pairs of consecutive numbers for each of the four rows. 
The pairs have been arranged so that the correspondence between fig.\ \ref{CalkinWilf} and 
fig.\ \ref{Pairs consecutive numbers}  is clear.

Other sections of Stern's paper record additional properties of the tree,  which we do not discuss here.  For
example, Stern discusses how often each number appears as a sum number.   

\subsection{The Stern-Brocot Tree of Rationals}\label{The Stern-Brocot Tree of Rationals}

\begin{figure}[h]
\includegraphics[width=\textwidth]{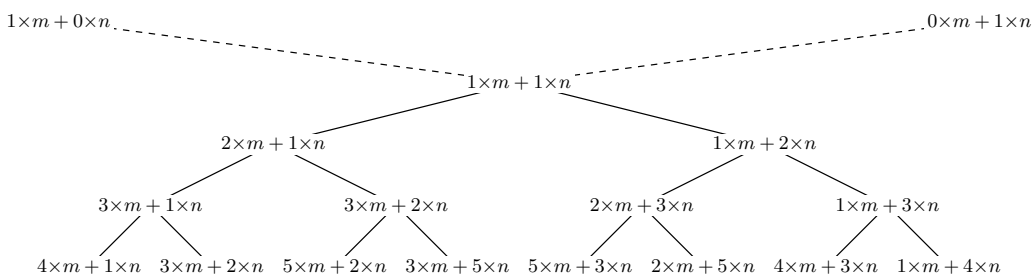} 
\caption{Tree of ``coefficients'' of  $\mathit{Ei\/}(m{,}n)$}\label{Stern coefficients}
\end{figure}

Identification of the so-called Stern-Brocot tree of rationals in Stern's paper is more demanding. 
Recall the process of constructing a sequence of rows of numbers from a given pair of numbers $m$ and $n$. 
It is clear that every number is a linear combination of $m$ and $n$.  Stern studies the \emph{coefficients} 
(``Coefficienten''), i.e.\ the pair of multiplicative factors of $m$ and $n$,  defined by the linear combination.  
Fig.\  \ref{Stern coefficients} displays the coefficients in a way that allows direct comparison with the 
Stern-Brocot tree
of  rationals (fig.  \ref{SternBrocot}).  (The reader may also wish to compare 
fig.\  \ref{Stern coefficients} with Graham, Knuth and Patashnik's depiction of the tree 
\cite[p. 117]{Concrete2}.) 

The numbers at the top-left and top-right of fig.\  \ref{Stern coefficients} are the numbers $m$ and $n$ written
as $1{\times}m\ms{1}{+}\ms{1}0{\times}n$ and $0{\times}m\ms{1}{+}\ms{1}1{\times}n$, respectively, in order to make the coefficients clear. This, we recall, is the
zeroth row in Stern's structure.  

In the subsequent levels of the tree, only the sum elements are displayed.  The correspondence between
fig.\ \ref{Stern coefficients} and fig.\ \ref{SternBrocot} should be easy to see; the number $k{\times}m\ms{1}{+}\ms{1}l{\times}n$ in fig.\ 
\ref{Stern coefficients} is displayed as the rational $\frac{l}{k}$ in fig.\ \ref{SternBrocot}.  The ``fundamental fact''  (4.31) in
\cite{Concrete2} is observed by Stern \cite[equation (8), p.207]{stern1858:rationals}  and used immediately to
infer that coefficients are relatively prime.   In section 15 of his paper, 
 Stern uses the (already proven) fact that the
Eisenstein-Stern tree is a tree of (all) rationals  to deduce that the Stern-Brocot tree is also a tree of
rationals.

\subsection{Newman's Algorithm}\label{Newman's Algorithm}

An interesting question is whether Stern also documents the algorithm currently attributed to Moshe
Newman for enumerating the elements of the Eisenstein array.  This is a question we found difficult to
answer because of our limited understanding of German.  However, the  answer would appear to be: 
almost, but not quite!

As remarked earlier, Stern documents a number of properties of groups of numbers in rows of
the Eisenstein
array, in particular groups of size three.  Of course, a group of size three comprises two groups of size
two.  Since groups of size two in the Eisenstein array  correspond to rationals in the Eisenstein-Stern
tree, by studying groups of size three Stern is effectively studying consecutive rationals in the
Eisenstein-Stern tree of rationals.  

It is important to note that Stern's focus is the sequence of \emph{rows} of numbers (in modern terminology, the
tree of numbers) as opposed to the (flattened) sequence of numbers defined by $\mathit{Ei\/}(m{,}n)$ --- 
significantly, the last number in
one row and the first number in the next row do not form a ``group'' according to Stern's definition.  This
means that, so far as we have been able to determine,  he nowhere considers a triple of numbers that
crosses a row boundary.  

Newman's algorithm (in the form we use  in section \ref{The Enumerations}) 
predicts that each triple of numbers in a given row of $\mathit{Ei\/}(1{,}1)$   has the form\begin{displaymath}
a\ms{5}b\ms{7}(2{}\left\lfloor{\displaystyle\frac{a}{b}}\right\rfloor\ms{1}{+}\ms{1}1)\ms{1}{\times}\ms{1}b\ms{1}{-}\ms{1}a 
\end{displaymath}
(Variable names have been chosen to facilitate comparison with Stern's paper.)  It follows immediately
that the sum of the  two outer elements of the triple is divisible by the middle element (that is,
$a\ms{1}{+}\ms{1}((2{}\left\lfloor{\textstyle\frac{a}{b}}\right\rfloor\ms{1}{+}\ms{1}1)\ms{1}{\times}\ms{1}b\ms{1}{-}\ms{1}a)$ is divisible by $b$); this fact is observed by Stern (for triples in a given row) in section
4 of his paper.  Importantly for what follows, Stern observes that the property holds  for $\mathit{Ei\/}(m{,}n)$ for
arbitrary natural numbers $m$ and $n$, and not just $\mathit{Ei\/}(1{,}1)$.   Stern  observes further \cite[(4) p.198]{stern1858:rationals} that each triple in
$\mathit{Ei\/}(m{,}n)$  has the form\begin{equation}\label{Sterntriple}
a\ms{5}b\ms{7}(2{}t\ms{1}{+}\ms{1}1)\ms{1}{\times}\ms{1}b\ms{1}{-}\ms{1}a 
\end{equation}
for some number $t$.     Stern identifies  $t$ as the number of rows preceding the current row in which the
number $b$ occurs as a sum element.  (In particular, if $b$ is a sum element then $t$ equals $0$.)  Stern shows how to
calculate  $t$ from the position of $b$ in  the row --- effectively by expressing the position as a binary
numeral.     (Note that ``$t$'' is the variable name used in Stern's paper; it has the same role as the variable
``$j$'' in our derivation of the algorithm in section \ref{The Enumerations}.)  

So far as we have been able to determine, Stern does not explicitly 
remark that  $t$ equals  $\left\lfloor{\textstyle\frac{a}{b}}\right\rfloor$ in the case of $\mathit{Ei\/}(1{,}1)$, but he does so implicitly in section 10 where he relates the
the continued fraction representation of ${\textstyle\frac{a}{b}}$ to the row number in which the pair $(a,b)$ occurs.
He does not appear to suggest a similar method for computing $t$ in
the general case of enumerating $\mathit{Ei\/}(m{,}n)$.  However, it is straightforward  to combine our derivation of 
 Newman's algorithm
with Stern's theorems to obtain an algorithm to enumerate the elements of $\mathit{Ei\/}(m{,}n)$ for arbitrary natural
numbers  $m$ and $n$.  Interested readers may consult our website \cite{jff*09:euclid-alg}
where several implementations are discussed.

As stated at the beginning of this section, the 
conclusion is  that Stern almost derives Newman's algorithm,
but not quite.   On the other hand, because his analysis is of the  general case $\mathit{Ei\/}(m{,}n)$ as opposed to $\mathit{Ei\/}(1{,}1)$,
his results are more general. 

\subsection{Stern-Brocot Enumeration}\label{Stern-Brocot Enumeration}

We now turn to the question whether Stern also gives an algorithm for enumerating the rationals in
Stern-Brocot order. 

To this end, we observe  that the form (\ref{Sterntriple}) extends to the coefficients  of each element 
of   $\mathit{Ei\/}(M{,}N)$, and hence to the elements of the Stern-Brocot tree.  Specifically, triples in $\mathit{Ei\/}(M{,}N)$ have the
form\begin{displaymath}
n_{0}{}M{+}m_{0}{}N\ms{7}n_{1}{}M{+}m_{1}{}N\ms{7}((2{}k\ms{1}{+}\ms{1}1){}n_{1}\ms{1}{-}\ms{1}n_{0}){}M\ms{1}{+}\ms{1}((2{}k\ms{1}{+}\ms{1}1){}m_{1}\ms{1}{-}\ms{1}m_{0}){}N 
\end{displaymath}
It is easy to exploit this formula directly to get an enumeration of the rationals in Stern-Brocot order, just
as we did above to obtain an enumeration of $\mathit{Ei\/}(M{,}N)$.  Just recall that the Stern-Brocot rationals are
given by the coefficients of the sum elements, and the sum elements are the odd-numbered elements in the
rows of $\mathit{Ei\/}(M{,}N)$ (where numbering starts from zero).  The algorithm so obtained is the one we derived in
section \ref{The Enumerations}.  

In this sense, Stern does indeed provide an  algorithm for enumerating the rationals in Stern-Brocot
order, albeit implicitly.  However, as with Newman's algorithm, he fails to observe the concise formula for 
the value of the variable $k.$
Also, a major methodological difference is our exploitation of the concision and precision afforded by matrix
algebra.  Given the state of development of matrix algebra in 1858,  Stern cannot be criticised  for not
doing the same.

Finally, we remark that  Stern returns to the properties of triples in section 19 of his paper.  
Unfortunately, we have
been unable to fully understand this section.  

\section{Brocot, the Watchmaker}\label{Brocot, the Watchmaker}

Achille Brocot was a famous French watchmaker who, some years before the publication of his paper
\cite{brocot-original}, 
had to fix some pendulums used for  astronomical measurements. However, the device was incomplete
and he did not know how to compute the number of teeth of  cogs that were missing. 
He was unable to  find any literature helpful to the solution of the problem, so, 
after some experiments, he devised a method to compute the  numbers.
In his paper, Brocot illustrates his method  with the following example:
\begin{quote}
A shaft turns once in 23 minutes. We want suitable cogs  so that another shaft completes a revolution
in 3 hours and 11 minutes, that is 191 minutes.
\end{quote}
The ratio between both speeds is $\frac{191}{23}$, so we can clearly choose a cog with $191$ teeth, and
another one with $23$ teeth. But, as Brocot wrote, it was not possible, at that time, to create cogs with so 
many teeth. And because $191$ and $23$ are coprime, cogs with fewer teeth can only approximate the 
true ratio.

Brocot's contribution was  a method to compute approximations to the true ratios (hence the title of his paper,
``Calculus of cogs by approximation''). He begins by observing that $\frac{191}{23}$ must be between the ratios  $\frac{8}{1}$  and
$\frac{9}{1}$. If we choose the ratio $\frac{8}{1}$, the error is ${-}7$ since $8{\times}23\ms{3}{=}\ms{3}1{\times}191\ms{2}{-}\ms{2}7$.  This means that if we choose this ratio,
the slower cog completes its revolution seven minutes early, i.e., after $8{\times}23$ minutes. 
On the other hand, if we choose the ratio $\frac{9}{1}$, the error is $16$ since $9{\times}23\ms{3}{=}\ms{3}1{\times}191\ms{1}{+}\ms{1}16$, meaning that the 
slower cog  completes its revolution sixteen minutes late, i.e., after $9{\times}23$ minutes.

Accordingly, Brocot writes two rows:

\begin{center}
$\begin{array}{p{2em}p{2em}p{2em}}
\MPnewitem 8 & 1 & {-}7\MPinsertsep 
\MPnewitem ~ & ~ & ~\MPinsertsep 
\MPnewitem 9 & 1 & {+}16\MPinsertsep 
\end{array}$
\end{center}
His method consists in iteratively forming a new row, by adding the numbers in all 
three columns of the rows that produce  the smallest error. 
Initially, we only have two rows, so we add the numbers in the three columns and we
write the row of sums in the middle.

\begin{center}
$\begin{array}{p{2em}p{2em}p{2em}}
\MPnewitem 8 & 1 & {-}7\MPinsertsep 
\MPnewitem 17 & 2 & {+}9\MPinsertsep 
\MPnewitem 9 & 1 & {+}16\MPinsertsep 
\end{array}$
\end{center}
(If we choose the ratio $\frac{17}{2}$, the slower cog completes its revolution $\frac{9}{2}$ minutes later, since $\frac{17}{2}\ms{2}{=}\ms{2}\frac{191{+}\frac{9}{2}}{23}$.)
Further approximations are constructed by adding a  row adjacent to the  row
that minimises the error term. The process ends once we reach the error $0$, which refers to the true ratio.
The final state of the table is:

\begin{center}
$\begin{array}{p{2em}p{2em}p{2em}}
\MPnewitem 8 & 1 & {-}7\MPinsertsep 
\MPnewitem 33 & 4 & {-}5\MPinsertsep 
\MPnewitem 58 & 7 & {-}3\MPinsertsep 
\MPnewitem 83 & 10 & {-}1\MPinsertsep 
\MPnewitem 191 & 23 & 0\MPinsertsep 
\MPnewitem 108 & 13 & {+}1\MPinsertsep 
\MPnewitem 25 & 3 & {+}2\MPinsertsep 
\MPnewitem 17 & 2 & {+}9\MPinsertsep 
\MPnewitem 9 & 1 & {+}16\MPinsertsep 
\end{array}$
\end{center}
The conclusion is that the two closest approximations to $\frac{191}{23}$ are ratios of  $\frac{83}{10}$  (which runs  $\frac{1}{10}$ minutes faster) 
and $\frac{108}{13}$ (which runs  $\frac{1}{13}$ minutes slower).  We could continue this process, getting at each stage
a closer approximation to $\frac{191}{23}$. In fact, Brocot refines the table shown above, in order to construct a multistage
cog train (see \cite[p. 191]{brocot-original}).

At each step in Brocot's process we add a new ratio  $\frac{m{+}m'}{n{+}n'}$, which is usually called the mediant of $\frac{m}{n}$ and
$\frac{m'}{n'}$.  Similarly,  each node in the Stern-Brocot  tree  is of the form $\frac{m{+}m'}{n{+}n'}$, where $\frac{m}{n}$ is the nearest ancestor
above and  to the left, and $\frac{m'}{n'}$ is the nearest ancestor above  and to the right. (Consider, for example, the
rational $\frac{4}{3}$ in figure  \ref{SternBrocot}. Its nearest ancestor above and to the left is $\frac{1}{1}$ and its nearest
ancestor above and to the right is $\frac{3}{2}$.) Brocot's process can be used to construct the  Stern-Brocot tree: first,
create an array that contains initially the  rationals $\frac{0}{1}$ and $\frac{1}{0}$; then, insert the rational $\frac{m{+}m'}{n{+}n'}$ between two
adjacent fractions $\frac{m}{n}$ and $\frac{m'}{n'}$.  In the first step we add only one rational to the array
\begin{displaymath}\frac{0}{1}\ms{1}{,}\ms{1}\frac{1}{1}\ms{1}{,}\ms{1}\frac{1}{0}\mbox{\ \ \ ,}\end{displaymath}but in the second step we add two new rationals:\begin{displaymath}\frac{0}{1}\ms{1}{,}\ms{1}\frac{1}{2}\ms{1}{,}\ms{1}\frac{1}{1}\ms{1}{,}\ms{1}\frac{2}{1}\ms{1}{,}\ms{1}\frac{1}{0}\mbox{\ \ \ .}\end{displaymath}Generally, in the $n^{\mathit{th\/}}$ step we add $2^{n{-}1}$ new rationals. Clearly, this array can be represented as an infinite
binary tree,  whose first four levels are represented in figure \ref{SternBrocot} (we omit the fractions $\frac{0}{1}$
and $\frac{1}{0}$).

The most interesting aspect to us of  Brocot's paper is that it solves an \emph{algorithmic} problem.  Brocot was
faced with the practical  problem of how to approximate rational numbers in order to construct clocks of
satisfactory accuracy and his solution is undisputably an \emph{algorithm}.   Stern's paper is closer to a traditional
mathematical paper but, even so, it is an in-depth study of an algorithm for generating rows of numbers
of increasing length. 

\section{Conclusion}\label{Conclusion}

There can be no doubt that what has been dubbed in recent years the ``Calkin-Wilf'' tree of rationals is, in
fact, a central topic in  Stern's 1858 paper.   
Calkin and Wilf  \cite{Calkin-Wilf2000}  admit that in Stern's paper
``there is a structure that is essentially our tree of fractions'' but add   ``in a different garb'' and 
do not clarify  what is meant by ``a different garb''.   It is unfortunate
that the misleading name has now become prevalent; in order to avoid further
misinterpretations  of historical fact, it would be desirable for Stern's  paper to be translated into English.

We have not attempted to determine how the name ``Stern-Brocot'' tree came into existence.  It has been
very surprising to us how much  easier  it is to identify the Eisenstein-Stern tree in Stern's paper in
comparison to identifying the Stern-Brocot tree.

\paragraph{Acknowledgements}\label{Acknowledgements}
Thanks go to Christian Wuthrich for help in translating the most relevant parts of Stern's paper,  
and to Jeremy Gibbons for his comments on earlier drafts of this paper, and for help with TeX
commands.    Thanks also to our colleagues in the Nottingham Tuesday Morning Club for helping iron
out omissions and ambiguities, and to Arjan Mooij and Jeremy Weissmann for 
their comments on section \ref{Distributivity properties}.

This work is (partially) funded by ERDF - European Regional Development Fund through the COMPETE Programme (operational programme for competitiveness) and by National Funds through the FCT - the Portuguese Foundation for Science and Technology, within project FCOMP-01-0124-FEDER-007254.

\end{document}